\documentclass[aps,prd,amsmath,preprint,longbibliography,superscriptaddress,nofootinbib]{revtex4-1}
\usepackage{geometry}
\usepackage[utf8]{inputenc}
\usepackage[T1]{fontenc}
\usepackage{verbatim}
\usepackage{amsmath,amssymb,amsfonts}	
\usepackage{graphicx}
\usepackage{color}
\usepackage{xcolor}

\usepackage{booktabs}

\usepackage{appendix}
\usepackage{slashed}
\usepackage{tikz}
\usepackage[compat=1.1.0]{tikz-feynhand}
\usetikzlibrary{shapes}
\usepackage{footmisc}
\usepackage[colorlinks=true,citecolor=blue,linkcolor=purple,urlcolor=purple]{hyperref}
\let\oldfootnote\footnote
\renewcommand{\footnote}[1]{%
    \begingroup%
    \linespread{1}%    % <- linespread for footnote: 1, 1.1, 1.2 etc
    \oldfootnote{#1}%
    \endgroup%
}

\makeatletter
\def\@ssect@ltx#1#2#3#4#5#6[#7]#8{%
  \def\H@svsec{\phantomsection}%
  \@tempskipa #5\relax
  \@ifdim{\@tempskipa>\z@}{%
    \begingroup
      \interlinepenalty \@M
      #6{%
       \@ifundefined{@hangfroms@#1}{\@hang@froms}{\csname @hangfroms@#1\endcsname}%
       {\hskip#3\relax\H@svsec}{#8}%
      }%
      \@@par
    \endgroup
    \@ifundefined{#1smark}{\@gobble}{\csname #1smark\endcsname}{#7}%
  }{%
    \def\@svsechd{%
      #6{%
       \@ifundefined{@runin@tos@#1}{\@runin@tos}{\csname @runin@tos@#1\endcsname}%
       {\hskip#3\relax\H@svsec}{#8}%
      }%
      \@ifundefined{#1smark}{\@gobble}{\csname #1smark\endcsname}{#7}%
      \addcontentsline{toc}{#1}{\protect\numberline{}#8}%
    }%
  }%
  \@xsect{#5}%
}%
\makeatother

\begin{document}
\preprint{ULB-TH/22-07}\preprint{CPPC-2022-04}
\heavyrulewidth=.08em
\lightrulewidth=.05em
\cmidrulewidth=.03em
\belowrulesep=.65ex
\belowbottomsep=0pt
\aboverulesep=.4ex
\abovetopsep=0pt
\cmidrulesep=\doublerulesep
\cmidrulekern=.5em
\defaultaddspace=.5em

\baselineskip 0.7cm

\bigskip

\title{Freeze-in and freeze-out of sterile neutrino dark matter}
\author{Rupert Coy}
\email[e-mail: ]{rupert.coy@ulb.be}
\affiliation{Service de Physique Théorique, Université Libre de Bruxelles,\\
Boulevard du Triomphe, CP225, 1050 Brussels, Belgium}
\author{Michael A.~Schmidt}
\email[e-mail: ]{m.schmidt@unsw.edu.au}
\affiliation{
Sydney Consortium for Particle Physics and Cosmology, School of Physics, The University of New South Wales,
Sydney, New South Wales 2052, Australia
}

\begin{abstract}
A sterile neutrino with a keV-scale mass is a compelling dark matter candidate. We propose a new production mechanism involving the decay and annihilation of a complex scalar singlet with a Higgs portal coupling which develops a vacuum expectation value. The interactions of the resulting pseudo Nambu-Goldstone boson may thermalise the dark sector.
We determine the region of parameter space where dark sector thermalisation is reached and discuss the most relevant cosmological observables.
The scenario can be considered as the combination of a freeze-in of the dark sector followed by relativistic freeze-out. 
\end{abstract}

\maketitle

\tableofcontents

\section{Introduction}
\label{intro}
%%%%%%%%%%%%%%%%%%%%%%%%%%%%%%%%%%%%%%%%%%%%%%%%%
Although the Standard Model (SM) is a very successful model of nature, it remains incomplete. 
Notably, both the nature of dark matter (DM), which accounts for one quarter of the energy density of the Universe~\cite{Planck:2018vyg}, and the origin of neutrino masses are unknown. 
The simplest explanation of the former is in terms of a new particle which is stable on cosmological timescales, interacts at most weakly with known baryonic matter, and is cold. 
Neutrinos themselves were one of the early dark matter candidates, but have long since been ruled out because their contribution to the Universe's energy density is much too small and their lightness contradicts our understanding of large scale structure formation.
An intriguing possibility for dark matter, however, is that it is comprised of sterile neutrinos with keV-scale mass~\cite{Dodelson:1993je}, thereby linking DM to the neutrino mass problem. 

Right-handed sterile neutrinos are a natural extension of the SM from a theoretical point of view, as they serve as the missing right-handed partners of the left-handed active neutrinos, moreover they can explain tiny neutrino masses through the seesaw mechanism \cite{Minkowski,Yanagida,GellMann,MohapatraSenj}. Three right-handed neutrinos are sufficient for simultaneously describing dark matter, via a single keV-scale sterile neutrino, and generating neutrino masses, via two heavy sterile neutrinos~\cite{Asaka:2005an}. 
Sterile neutrinos with a keV-scale mass are generally warm dark matter candidates (WDM) which suppress power at small scales and therefore may play some role in resolving small scale structure issues, although explanations involving baryonic physics or self-interacting DM have been more favoured recently \cite{Bullock:2017xww}. 
Sterile neutrino DM with 7.1 keV mass has also been suggested as a possible explanation of the 3.55 keV x-ray line \cite{Bulbul:2014sua,Boyarsky:2014jta}. 
More generally, the keV scale is an intriguing frontier to consider because both the Tremaine-Gunn bound \cite{Tremaine:1979we} and Lyman-$\alpha$ data \cite{Horiuchi:2013noa} forbid fermionic DM from being lighter than $\mathcal{O}$(keV).

Such light DM cannot be frozen out by electroweak interactions, because in that case it would be overabundant~\cite{Lee:1977ua}. 
A simple, viable production mechanism for keV-scale DM is freeze-in~\cite{McDonald:2001vt,Hall:2009bx}, where its abundance is initially negligible and is slowly produced via decay, annihilation or mixing controlled by a small coupling. 
In the presence of active-sterile neutrino mixing, sterile neutrinos are most straightforwardly produced via neutrino oscillations~\cite{Barbieri:1990vx,Enqvist:1990ad,Dodelson:1993je}. However, non-resonant neutrino oscillations has already been ruled out as the source of the overall DM abundance due to a combination of x-ray and Ly-$\alpha$ constraints~\cite{Boyarsky:2008xj}. Even in presence of a large lepton asymmetry, resonant neutrino  oscillations~\cite{Shi:1998km} are still strongly restricted by Ly-$\alpha$ data~\cite{Horiuchi:2013noa}. 

A minimal extension of this scenario involves the production of sterile neutrinos from scalar boson decay~\cite{Shaposhnikov:2006xi,Kusenko:2006rh,Petraki:2007gq,Boyanovsky:2008nc,Merle:2013gea,Merle:2013wta,Frigerio:2014ifa,Adulpravitchai:2014xna,Kang:2014cia,Merle:2015oja,Adulpravitchai:2015mna,DeRomeri:2020wng}. 
Within this, we will consider the possibility that the scalar is itself frozen-in~\cite{Merle:2013wta,Adulpravitchai:2014xna,Kang:2014cia,Merle:2015oja,DeRomeri:2020wng}, thus production is a two-step process. Initially, the scalar boson is produced; subsequently, it decays to sterile neutrinos. 
Existing studies have focused only on a real scalar field which is odd under a $\mathbb{Z}_2$ symmetry. After spontaneous symmetry breaking of the discrete symmetry, the scalar boson develops a vacuum expectation value and generates the sterile neutrino mass. Thus, the coupling controlling the scalar decay to sterile neutrinos is directly proportional to the sterile neutrino mass. The interactions of the sterile neutrinos (and often of the frozen-in scalar) are feeble enough that they do not thermalise. 
However, the breaking of the global discrete $\mathbb{Z}_2$ symmetry can lead to the formation of domain walls~\cite{Zeldovich:1974uw}, which is an intrinsic problem of this real scalar singlet model.

In this work we propose an elegant way to avoid domain walls by promoting the discrete $\mathbb{Z}_2$ symmetry to a global $U(1)$ symmetry, and study its phenomenological implications. 
It requires the introduction of a complex scalar field, rather than a real scalar, to generate a sterile neutrino mass term after spontaneous symmetry breaking. This small change drastically modifies the phenomenology of sterile neutrino production in the early Universe. 
Interactions involving the pseudo Nambu-Goldstone boson (pNGB) 
can enable the frozen-in scalar sector to reach both chemical and kinetic equilibrium and thus to thermalise with a temperature $T_\phi< T$ (the model is therefore a concrete example of thermal DM with $T_\phi \neq T$, which was recently classified in~\cite{Coy:2021ann}). 
It may even bring the sterile neutrinos into kinetic (and chemical) equilibrium. 
Since the interactions eventually go out of equilibrium and the sterile neutrino comoving number density becomes fixed at some temperature below the heavy scalar mass, it can be said to freeze-in and freeze-out.

This scenario is a departure from the usual case, as generally neither kinetic nor chemical equilibrium is reached for frozen-in fields which explain the observed dark matter abundance. 
In this work, we also include quantum statistical factors and apply the full relativistic treatment, making approximations where applicable. 
This is important in the context of the light sterile neutrino and pNGB.

The paper is organised as follows. In Sec.~\ref{sec:model} we outline the production mechanism in general terms and introduce the model. Sterile neutrino production is discussed in Sec.~\ref{sec:Nproduction}, and the temperature of the dark sector is computed in Sec.~\ref{CalcTphi}. 
The thermalisation of the scalar sector is addressed in Sec.~\ref{sec:thermalisationPhi}, and the thermalisation of the sterile neutrinos in Sec.~\ref{sec:thermalisationN}. Finally, we discuss the results and conclude in Sec.~\ref{sec:conclusion}. In the appendices we present some technical details related to scattering in the scalar sector.
%%%%%%%%%%%%%%%%%%%%%%%%%%%%%%%%%%%%%%%%%%%%%%%%%%%%%%%%%%%%%%%%%%%%%%%%%%%%%%%%%%%%%%%%%%%%%%

\section{The model}
\label{sec:model}

The freeze-in mechanism \cite{McDonald:2001vt,Hall:2009bx} is now a well-known scenario for the production of particles whose initial abundance is assumed to be negligible, possibly due to the inflationary dynamics.\footnote{The impact of a non-negligible initial dark sector abundance was recently studied in \cite{Fernandez:2021iti}.} 
A key feature of this mechanism is that interactions between the frozen-in particles and the thermal bath of SM particles (which has temperature $T$) is sufficiently feeble that the former never equilibrate with the latter. 
This feeble interaction is due to a very small coupling between the two sectors, which can nevertheless be technically natural \cite{tHooft:1979rat}. 
The production predominantly takes place at temperatures of the order of the masses of the parent particles, as long as the interactions are renormalisable. 

\begin{figure}[tb!]\centering
	\begin{tikzpicture}
		\node[ellipse,red,inner sep=0.65cm,draw,line width=1mm] (SM) {\large SM};
		\node[below,yshift=-2ex] at (SM) {\large $T$};
		\node[ellipse,blue,inner sep=0.65cm,draw,right,shift={(2.5,0)},line width=1mm] (phi) at (SM.east) {\large $\varphi, \alpha$};
		\node[below,yshift=-2ex] at (phi) {\large $T_\phi$};
		\node[ellipse,blue,inner sep=0.65cm,draw,right,shift={(2.5,0)},line width=1mm] (N) at (phi.east) {\large $N$};
		\node[below,yshift=-2ex] at (N) {\large ($T_\phi$)};
		\draw[->,dashed,line width=1mm] (SM) -- node[midway,below=1ex,sloped,align=center] {$h,hh\rightarrow \varphi,\varphi\varphi$\\$WW,ZZ,t\bar t \to \varphi\varphi$} (phi);
		\draw[<->,solid,line width=1mm] (phi) -- node[midway,below=1ex,sloped] {$\varphi,\alpha\alpha\leftrightarrow NN$} 
		(N);
	\end{tikzpicture}
	\caption{Different sectors and the relevant interactions. The SM-$\phi$ interactions do not thermalise, but the $\phi$ sector self-scattering leads to thermal equilibrium such that the hidden sector develops a temperature $T_\phi<T$. The $\phi$-N interactions may or may not thermalise the $N$ population to the same temperature $T_\phi$. 
	}
	\label{fig:sectors}
	\end{figure}
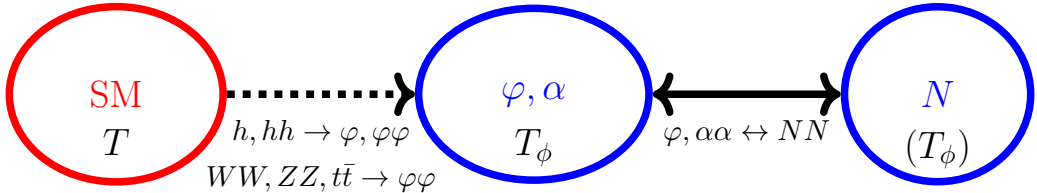

The mechanism we consider in this paper is schematically illustrated in Fig.~\ref{fig:sectors}. It involves a second iteration of the freeze-in scenario.\footnote{The scenario of successive freeze-in processes, i.e.~$A \to B$ via freeze-in and then $B \to C$ also via freeze-in, where $C$ is the DM, is called `sequential freeze-in' in \cite{Hambye:2019dwd}, see their Sec.~VI for a related model.}  
A complex scalar $\phi$ (which consists of a heavy scalar $\varphi$ and pNGB $\alpha$, see Eqn.~\eqref{Higgs}) and sterile neutrino $N$ both have negligible initial abundance. 
First, $\varphi$ is frozen-in via the decay or annihilation of particles in the thermal plasma. 
The $\phi$ itself equilibrates through self-interactions, forming its own thermal bath of $\varphi$ and $\alpha$ with some temperature $T_\phi < T$. 
Note that models with only a real scalar typically do not thermalise~\cite{Drewes:2016upu,Boyarsky:2018tvu}, although there are exceptions, e.g.~\cite{DeRomeri:2020wng}. 
The presence of the pNGB component of the complex scalar enables rapid kinetic and chemical equilibration, which is a distinguishing feature between models with a real scalar and a complex one. 
In this work, we entirely focus on the region in parameter space where thermalisation is obtained. The relevant conditions for achieving thermalisation for the dark scalar sector are discussed in Secs.~\ref{sec:thermalisationPhi} and \ref{sec:thermalisationN}.
We may describe $\phi$ in terms of its components $\varphi$ and $\alpha$ as long as $T_\phi \lesssim T_{\phi,c}$, the critical temperature for the dark sector phase transition. Before the dark sector phase transition, the scalar is described as one complex scalar field $\phi$.

\begin{figure}
    \centering
    \includegraphics[width=0.8\textwidth]{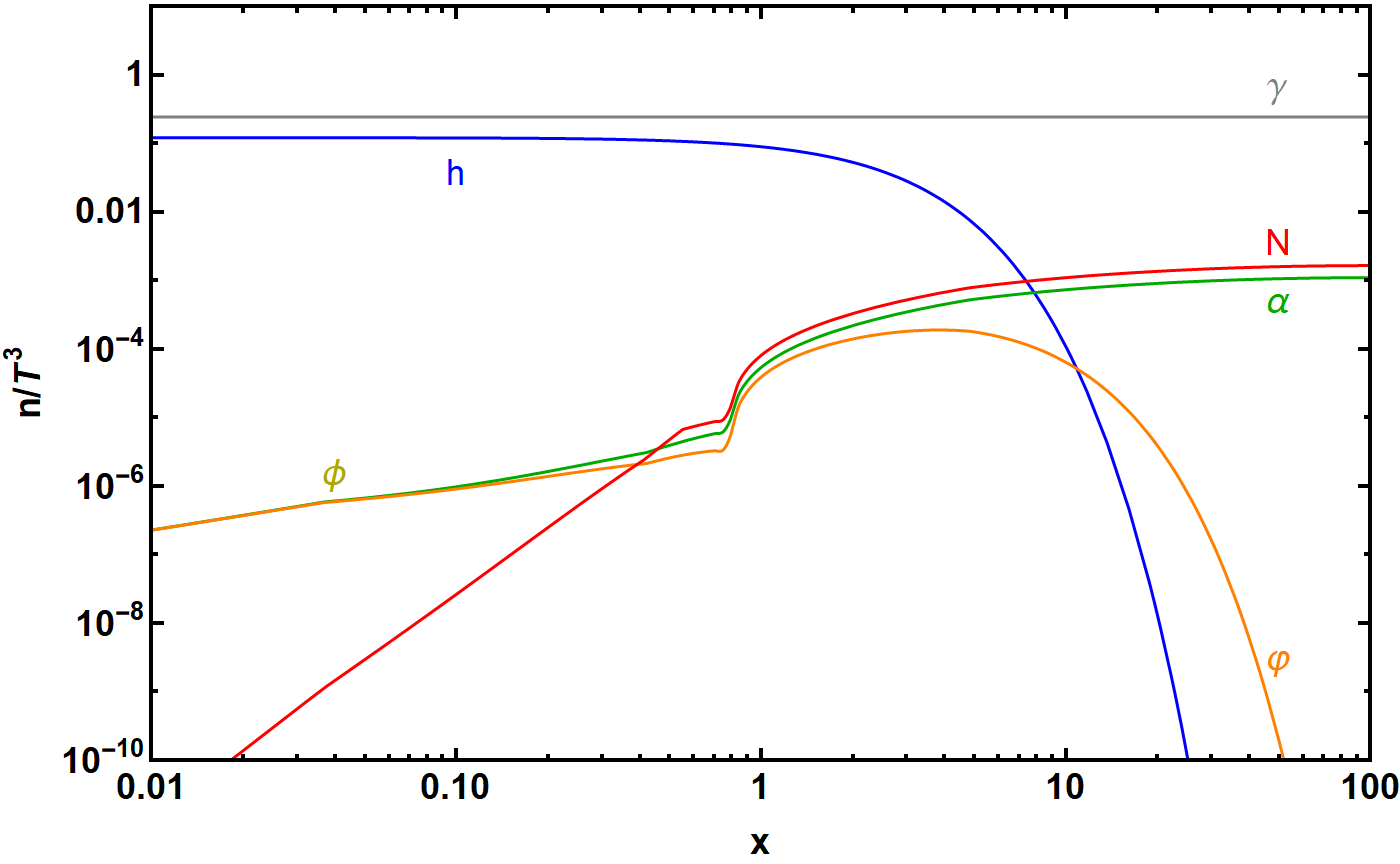}
    \caption{Schematic plot of the particle number evolution, in terms of $x = m_h/T$. Here $m_\varphi = 10$ GeV, $m_N = 10$ keV, $\lambda_\phi = 1$, while $\kappa = 5.3 \times 10^{-9}$ is fixed to ensure that sterile neutrinos give the correct DM relic abundance. 
    The grey, blue, red, green and orange lines correspond to the photon, Higgs, sterile neutrino, $\alpha$ and $\varphi$ number densities rescaled by $T^3$. Early on $T_\phi > T_{\phi ,c}$, thus there is a single number density of the complex scalar $\phi$. 
    The rapid increase in the $N$, $\alpha$ and $\varphi$ number densities at $x\simeq 0.8$ is due to $\varphi$ production from Higgs decays after the EW phase transition, see the text.}
    \label{fig:evolplot}
\end{figure}

When $T_\phi \sim m_\varphi$, the $N$ population freezes in mainly from decays of the scalar $\varphi$, whose population then becomes Boltzmann suppressed, and annihilations of the $\alpha$. 
Here, we consider the case where the $N$ population eventually thermalises with the $\phi$ bath and thus is also described by temperature $T_\phi$. However, we note that it is possible that the $N$ population never thermalises, as we discuss in Sec.~\ref{sec:thermalisationN}. 
The behaviour of the different sectors is outlined in Fig.~\ref{fig:evolplot}.

In our explicit realisation of the mechanism we introduce in the theory a complex scalar singlet, $\phi$, and a light right-handed singlet fermion, $N_R$, with lepton number -2 and 1, respectively (the SM lepton doublet and singlet, $L$ and $e$, have lepton number 1). 
The scalar and leptonic parts of the Lagrangian are
\begin{align}
\label{lagrangian}
	\mathcal{L} &= 
	(D_\mu H)^\dagger (D^\mu H)
+ \partial_\mu \phi^\dagger \partial^\mu \phi 
-  V(H,\phi)
	+ i\overline{N_R} \slashed{\partial} N_R
	+ i\overline{L} \slashed{D} L
	+ i\overline{e_R} \slashed{D} e_R
\\\nonumber&
	- y^e_{i}\left( \overline{L_i} H e_{iR} + \mathrm{h.c.}\right)
	- \left(c_{ij} (\overline{L_i^c} \epsilon H) (L_j \epsilon H) + \mathrm{h.c.} \right)
	- \frac{f}{2} \left(\phi N_R^T C N_R + \mathrm{h.c.}\right)
	- \left(y^\nu_i\,  \overline{L_i}\tilde H N_R + \mathrm{h.c.}\right) 
	\;,
\end{align}
where $C$ denotes the charge conjugation matrix, $\epsilon=i\sigma_2$, $\tilde H= \epsilon H^*$, $L^c = C \overline{L}^T$ and the covariant derivative is defined as $D_\mu = \partial_\mu - i g_1 Y B_\mu - i g_2 W_\mu^a \sigma^a/2$. 
Light neutrino masses come from the Weinberg operator in the second line, which is e.g.~induced by heavier sterile neutrinos. 
The scalar potential is given by
\begin{equation}
V(H,\phi) = 
\lambda\left(H^\dagger H - \frac{v^2}{2}\right)^2
+ \lambda_\phi \left(|\phi|^2 -\frac{v_\phi^2}{2}\right)^2
+ \kappa \left(|\phi|^2-\frac{v_\phi^2}{2}\right) \left(H^\dagger H-\frac{v^2}{2}\right)
\;,
\label{Potential}
\end{equation}
and the scalar $\phi$ and Higgs doublet $H$ are decomposed as follows:\footnote{We decomposed the complex scalar, $\phi$, into modulus and phase, rather than real and imaginary parts. This simplifies calculations and makes the proportionality of matrix elements to the momenta of pNGBs $\alpha$ in the initial and final states explicit.}
\begin{align}
	\phi &= \frac{v_\phi + \varphi}{\sqrt{2}} e^{i \alpha/v_\phi}  &
	H & = \begin{pmatrix}
G^+ \\
\frac{v + h+i G^0}{\sqrt{2}}
\end{pmatrix} ,
\label{Higgs}
\end{align}
with the electroweak VEV $v = (\sqrt{2}G_F)^{-1/2}\simeq 246$ GeV. 
Without loss of generality we take charged-lepton Yukawa couplings $y^e_i$ real and diagonal. 
The coupling $f$ can be made real by a phase redefinition of $N$, the sterile neutrino field. 
The neutrino Yukawa couplings, $y^\nu_i$, are generally complex in the presence of massive neutrinos and naturally small, $|y^\nu_i| \ll 1$, because the symmetry of the model is enhanced in their absence (in conjunction with $|\kappa| \ll 1$). 
Specifically, decoupled $N_R$ and $\phi$ may transform under their own independent $U(1)$ symmetry with charges $1$ and $-2$, respectively. 

In Eqn.~\eqref{lagrangian} we take $\kappa \ll 1$, its smallness is protected by an enhanced Poincar\'e symmetry in the limit that $\kappa \to 0$ (see \cite{Foot:2013hna}). 
The potential at zero temperature is minimised by the VEVs, $v$ and $v_\phi$. 
The VEV of $\phi$ spontaneously breaks lepton number, 
giving mass to the sterile neutrino. 
As CP is conserved in the scalar sector, there is no mixing between the CP-even states $\varphi$, $h$ and the CP-odd pNGB $\alpha$, a Majoron~\cite{Chikashige:1980ui,Schechter:1981cv}.
The mass matrix for $h, \varphi$ is
\begin{equation}
M^2 = \begin{pmatrix}
2 \lambda v^2 & \kappa v v_\phi \\
\kappa v v_\phi & 2 \lambda_\phi v_\phi^2
\end{pmatrix}.
\label{MassMatrix}
\end{equation}
The mixing of $\varphi$ with the Higgs induces a coupling of the SM Higgs with two $\alpha$ particles and of $\varphi$ with a pair of SM fermions, however this is suppressed by the small mixing. 
Since the off-diagonal elements are tiny due to the factor of $\kappa$, we have 
		\begin{equation}
			m_h^2 \approx 2 \lambda v^2\quad \mathrm{and}\quad
			m_\varphi^2 \approx 2 \lambda_\phi v_\phi^2\;.
		\end{equation}
	The first relation gives $\lambda \approx 0.129$ in order to explain the SM Higgs boson mass of $m_h=125$ GeV. 
The CP-even scalar mass eigenstates $(s_1,s_2)$ are 
\begin{equation}
	\begin{pmatrix}
		h\\
		\varphi
	\end{pmatrix} =
	\begin{pmatrix}
		\cos\theta & -\sin\theta\\
		\sin\theta & \cos\theta\\
	\end{pmatrix}
	\begin{pmatrix}
		s_1\\s_2
	\end{pmatrix}~,
\end{equation}
with the mixing angle given by
\begin{equation}
\tan(2\theta) = \frac{v v_\phi \kappa}{\lambda v^2 - \lambda_\phi v_\phi^2} \approx \frac{2 v v_\phi \kappa}{m_h^2 - m_\varphi^2}~.
\label{Tan2Theta}
\end{equation}
The last approximation holds as long as $m_h^2-m_\varphi^2 \gg 4\kappa v
v_\phi$, but breaks down when the scalar masses are almost degenerate. 
As the mixing angle is very small, we will neglect it except when it is important for a physical process like the production of the new scalars from Higgs decay or annihilation. 
In this limit, the relevant interactions 
of the CP-even scalars are
\begin{align}
    V 
    & \supset
    \lambda v s_1^3
    +\lambda_\phi v_\phi s_2^3
    + \kappa \frac{s_1^2s_2^2}{4} 
    + \kappa\, v\,  \frac{2m_\varphi^2+m_h^2}{m_h^2-m_\varphi^2} \frac{s_1s_2^2}{2}
    - \kappa\, v_\phi\, \frac{2m_h^2 + m_\varphi^2}{m_h^2-m_\varphi^2} \frac{s_1^2 s_2}{2}\;. 
    \label{potentialMassBasis}
 \end{align}
In the following, we will simplify the notation and denote the mass eigenstates by the interaction eigenstate with the largest contribution, i.e. 
\begin{equation}
	s_1\approx h \quad \mathrm{and}\quad s_2 \approx \varphi .
\end{equation}
The spontaneous breaking of lepton number by the VEV of the $\phi$ leads to a pNGB, so $\alpha$ is (almost) massless even in the presence of thermal corrections due to the Goldstone theorem. 
Lepton number will generally be explicitly broken by quantum gravity effects and induce a tiny mass for the pNGB, which however we neglect in the following. 
Equally, the population of pNGBs produced in the early Universe via the misalignment mechanism is negligible in this
scenario~\cite{Marsh:2015xka}.
The VEVs depend on thermal corrections and become non-zero below the critical temperatures $T_{\rm ew,c}=159.5\pm1.5$ GeV~\cite{DOnofrio:2015gop} in the SM sector and $T_{\phi,c}=\sqrt{3 /(2\lambda_\phi)}m_\varphi$ in the dark sector. The dark sector generally has a lower temperature, $T_\phi$, than the SM. We parameterise it in terms of the temperature ratio
\begin{equation}
    z\equiv \frac{T_\phi}{T}\;.
\end{equation}
Although there is generally a period
in the cosmological evolution with $T_\phi>T_{\phi,c}$, it can often be neglected as long as the dominant production of dark sector particles occurs for $T\lesssim 12 m_\varphi (0.1/z)/\sqrt{\lambda_\phi}$, with the temperature ratio $z\lesssim 0.1$ required for the correct DM abundance. 
In the following, we will approximate the VEV evolution with step-functions and neglect the intermediate unbroken dark sector phase. We will justify this approximation at the relevant places of our analysis.

After the breaking of lepton number, we perform a local baryon minus lepton number transformation, 
\begin{align}
	(N_R,L,e_R) &\to e^{-i\alpha(x)/2v_\phi} (N_R,L,e_R) \\\nonumber
	(Q_L,u_R,d_R) & \to e^{i\alpha(x)/6v_\phi} (Q_L,u_R,d_R)
\end{align}
to make the derivative couplings of the pNGB explicit 
and to express all fields in the physical mass basis.
In particular, we introduce neutrino mass eigenstates $n_i$ with masses $m_i$ in terms of their flavour eigenstates,
\begin{equation}
    n_\alpha^\prime \equiv \begin{pmatrix}
        \nu_i \\ N_R^c \end{pmatrix}_\alpha = U_{\alpha i} n_i \; ,
\end{equation}
where the unitary matrix $U$ diagonalises the neutrino mass matrix
\begin{equation}
    M = \begin{pmatrix}
        m_\nu & m_{D} \\
        m_{D}^T & m_N
    \end{pmatrix}
    \qquad\qquad \mathrm{with}
    \qquad\qquad
    U^T M U = \mathrm{diag}(m_1,\dots, m_4)\;,
\end{equation}
$m_{D}= y^\nu v/\sqrt{2}$, $m_\nu=c v^2$ and $m_N = f v_\phi/\sqrt{2}$.
By construction, charged leptons are simultaneously mass and interaction eigenstates. Finally, we introduce Majorana neutrino fields $n_{Mi} = n_i + n_i^c$, and write the relevant part of the interaction Lagrangian for leptons and scalars as\footnote{Note that the transformation of the path integral measure does not induce couplings of the pNGB to field strength tensors, since the anomaly conditions linear in baryon minus lepton number and quadratic in the SM gauge groups are satisfied. However, the anomaly condition linear in baryon minus lepton number with a quadratic gravitational part and the cubic anomaly condition for baryon minus lepton number are not satisfied. They can be satisfied by introducing two additional sterile neutrinos.}
\begin{align}
	\mathcal{L} =& 
\left( \frac{\varphi}{v_\phi} + \frac{\varphi^2}{2v_\phi^2}\right) \partial_\mu \alpha\partial^\mu \alpha
	-  V(h,\varphi) 
	+ e A_\mu J^\mu_{\rm em} 
	+g \,\left(W_\mu^+ J_W^{\mu+}  + W_\mu^-J_W^{\mu-} + Z_\mu J_Z^\mu\right)
	 \\\nonumber&
 - \frac{m_N}{2 } \frac{\varphi}{v_\phi} \mathrm{Re}(U_{Ni} U_{Nj}) \,\overline{n_{Mi}} n_{Mj} 
 + \frac{m_N}{2 } \frac{\varphi}{v_\phi} \mathrm{Im}(U_{Ni} U_{Nj})\, \overline{n_{Mi}} i \gamma_5 n_{Mj} 
-  \frac{\partial_\mu \alpha}{4 v_\phi}\, \overline{n_{Mi}} \gamma^\mu \gamma_5  n_{Mj} (U^\dagger P U)_{ij} 
\end{align}
with the diagonal matrix $P=\mathrm{diag}(1,1,1,-1)$ which accounts for the different chiralities of the three left-handed neutrinos $\nu_i$ and the sterile neutrino $N_R$.
The electromagnetic and charged and neutral weak currents are:
\begin{align}
	J_{\rm em}^\mu & = -\overline{e_i}\gamma^\mu e_i\;,
	&
	J_W^{\mu - } & = \frac{U_{ij}}{\sqrt{2}} \,\overline{e_i} \gamma^\mu P_L n_{Mj}
	= J_W^{\mu+\dagger}\;,
	\end{align}
	\begin{align}
	J_Z^\mu & = \frac{1}{\cos\theta_w} \left\{ 
  \frac12\sum_{\alpha=1}^3 U_{\alpha i}^* U_{\alpha j}   \overline{n_{Mi}}\gamma^\mu \left(\frac12\right)n_{Mj}
+\overline{e}\gamma^\mu \left[ 
	\left(-\frac12 + \sin^2 \theta_w\right) P_L 
+ \left(\sin^2 \theta_w\right) P_R \right] e
	\right\}
	\;.
\end{align}

The pNGB $\alpha$ only has derivative couplings and does not couple to charged fermions,
because charged fermions couple as a vector current to the derivative of the pNGB. The derivative couplings of the pNGB to neutrinos 
mediate neutrino-neutrino scattering, which is most strongly constrained by cosmic microwave background (CMB) data from Planck to be $m_\nu/2v_\phi \lesssim 7\times 10^{-7}$ at 95\% CL~\cite{Forastieri:2019cuf,Berryman:2022hds}.
See~\cite{Berryman:2022hds} for a discussion of other constraints. The CMB constraint translates in a lower bound on the VEV of $v_\phi \gtrsim 0.7\,\mathrm{MeV}\, (m_\nu/1\mathrm{eV})$, which does not pose any constraint on the parameter space considered for the keV sterile neutrino DM production in the following. 

As the coupling of the pNGB $\alpha$ to neutrinos is flavour-violating, it will induce fast decays $N\to \nu \alpha$ proportional to the square of the active-sterile mixing, $U_{Ni} \simeq y^\nu_i v / (f v_\phi)$,  
\begin{equation}
  \Gamma(N\to \nu \alpha) \simeq \frac{m_N^3}{16\pi v_\phi^2} \sum_i |U_{Ni}|^2 \;.
\end{equation}
Requiring that the keV sterile neutrino lifetime exceeds the age of the universe, $\Gamma_N^{-1}> t_{\rm universe} = 4.4\times 10^{17}$ sec~\cite{Planck:2018vyg}, places a constraint on the active-sterile mixing of $\sum_i |U_{Ni}|^2< 7.6\times 10^{-26} \, (10\,\mathrm{keV}/m_N)^3 (v_\phi/\mathrm{GeV})$. There are additional contributions to keV sterile neutrino decay which are mediated by the flavour-violating interactions of the scalar $\varphi$ and the electroweak gauge bosons $Z$ and $W$.
The scalar $\varphi$ and in particular the $Z$ boson mediate tree-level decays to three light active neutrinos. 
Since the scalar $\varphi$ dominantly couples to the keV sterile neutrino, its contribution to the decay width is proportional to the sixth power of the active-sterile mixing and is therefore highly suppressed. Hence, tree-level sterile neutrino decay to three neutrinos is dominantly mediated by the $Z$ boson, and its decay width is~\cite{Pal:1981rm,Barger:1995ty}
\begin{equation}
\Gamma(N\to 3\nu) \simeq \frac{G_F^2 m_N^5}{96\pi^3} \sum_i |U_{Ni}|^2 \simeq 4.6\times 10^{-33}\, \mathrm{keV}\, \left(\frac{m_N}{10 \,\mathrm{keV}} \right)^5\sum_i |U_{Ni}|^2 \, ,
\end{equation}
where $G_F$ denotes the Fermi constant. Again demanding that the keV sterile neutrino lifetime exceeds the age of the universe, $\Gamma_N^{-1}> t_{\rm universe} = 4.4\times 10^{17}$ sec~\cite{Planck:2018vyg}, results in a constraint on 
$\sum_i |U_{Ni}|^2< 3.3\times 10^{-4}\, (10\,\mathrm{keV}/m_N)^5$~\cite{Boyarsky:2018tvu}. The scalar $\varphi$ also mediates decays to a light neutrino and two pNGBs, with 
\begin{equation}
    \Gamma(N \to \nu \alpha\alpha) \simeq \frac{\lambda_\phi^2 m_N^9}{7680 \pi^3 m_\varphi^8} \sum_i |U_{Ni}|^2 \simeq 4.2 \times 10^{-45} \,\mathrm{keV} \,\lambda_\phi^2\, \left( \frac{m_N}{10\, \mathrm{keV}}\right)^9 \left(\frac{1\,\mathrm{GeV}}{m_\varphi}\right)^8 \sum_i |U_{Ni}|^2 \, ,
\end{equation}
which is however suppressed by the phase space compared to $N\to \nu \alpha$. In summary, the lifetime of the keV sterile neutrino is set by the decay to an active neutrino and the pNGB, $N\to \nu \alpha$ and the other decay channels can be neglected.

At 1-loop order, charged-current interactions mediated by the $W$ boson induce radiative decay of the keV sterile neutrino to a light active neutrino and a photon, with~\cite{Pal:1981rm,Barger:1995ty}
\begin{equation}
    \Gamma(N\to \nu\gamma) \simeq \frac{9 \alpha_{\rm em} G_F^2 m_N^5}{256\pi^4} \sum_i|U_{Ni}|^2
    \simeq 3.6\times 10^{-35} \,\mathrm{keV} \left(\frac{m_N}{10\,\mathrm{keV}}\right)^5 \sum_i |U_{Ni}|^2 \, ,
\end{equation}
where $\alpha_{\rm em}$ denotes the fine structure constant.
This provides the second most stringent constraint on the active-sterile neutrino mixing, which is however much weaker than the lifetime constraint from $N\to \nu\alpha$.
For instance, for $m_N = 7.1$ keV, x-ray bounds constrain $|y^\nu_i|\lesssim 10^{-13}$. See \cite{Drewes:2016upu,Boyarsky:2018tvu} for a detailed discussion of the phenomenology of keV sterile neutrino DM. 
Hence, the production of the keV sterile neutrino via neutrino oscillations from active neutrinos is strongly constrained including non-resonant oscillations, which require 
a large lepton asymmetry~\cite{Shi:1998km}, which is not generated in this model. Hence, the contribution of active-sterile neutrino oscillations is subdominant and cannot account for the observed DM abundance. 

As the focus of this work is keV sterile neutrino production via Higgs decay, we neglect the active-sterile mixing in the following and thus the unitary matrix $U$ takes a block-diagonal form consisting of the Pontecorvo-Maki-Nakagawa-Sakata (PMNS) mixing matrix and $U_{N4}=1$. Using this approximation, the couplings of the pNGB are flavour-universal and the relevant terms in the interaction Lagrangian for keV sterile neutrino production in the early universe are
\begin{align}
	\mathcal{L} \supset& 
\left( \frac{\varphi}{v_\phi} + \frac{\varphi^2}{2v_\phi^2}\right) \partial_\mu \alpha\partial^\mu \alpha
	-  V(h,\varphi) 
 - \frac{m_N}{2 } \frac{\varphi}{v_\phi} \,\overline{N} N 
	 -  \frac{\partial_\mu \alpha}{4 v_\phi}\, \overline{N} \gamma^\mu \gamma_5 N
	 \;,
\end{align}
where we denote the keV sterile neutrino by $N \equiv n_{M4}$. 
In total, there are four independent parameters relevant for sterile neutrino production in the early Universe, 
the four what we work with being the Higgs-portal coupling $\kappa$, the quartic coupling $\lambda_\phi$, the $\varphi$ scalar mass $m_\varphi$, and the keV sterile neutrino mass $m_N$. 
The VEV, $v_\phi$, and the Yukawa coupling, $f$, can be expressed in terms of the other parameters and are given by
\begin{align}
	v_\phi & = \frac{m_\varphi}{\sqrt{2\lambda_\phi}} \, ,&
	f & = \frac{\sqrt{2} m_N}{v_\phi} = \frac{2 \sqrt{\lambda_\phi} m_N}{m_\varphi}\;. 
\end{align}

%%%%%%%%%%%%%%%%%%%%%%%%%%%%%%%%%%%%

\section{Sterile neutrino production and free-streaming horizon}
\label{sec:Nproduction}

As outlined in the previous section, the $\phi$ sector is frozen-in via Higgs portal interactions. The DM candidate, $N$, is then mainly produced via $\varphi \to NN$ decays and $\alpha \alpha \to NN$ annihilations, with a rate determined by the coupling $f \ll 1$. 
Scalar-pNGB scattering, $\alpha\varphi\to NN$, is suppressed compared to pNGB scattering for $T_\phi\ll m_\phi$, i.e.~when $\varphi$ is non-relativistic, while other processes are completely negligible since they are suppressed by additional powers of $f$. 
For large $T_\phi \gtrsim m_\varphi$, the decay and annihilation rates are comparable, however in the regime $T_\phi \ll m_\varphi$ the sterile neutrinos are dominantly produced via the annihilation process since the $\varphi$ population is exponentially suppressed. 

Assuming that the dark sector particles $\varphi$, $\alpha$ and $N$ are in thermal equilibrium (an assumption we will discuss in detail in Secs.~\ref{sec:thermalisationPhi} and \ref{sec:thermalisationN}), and neglecting inverse processes from the $\phi-N$ bath to the SM,\footnote{This is motivated by the standard freeze-in assumption that the $\phi$ and $N$ number densities are initially negligible \cite{McDonald:2001vt,Hall:2009bx}. Even at late times, after a substantial population of $\phi$ or $N$ has been frozen-in, we may continue to neglect these reverse processes because the $\phi$ and $N$ abundances are much smaller than the populations of their parents since $T_\phi \ll T$.} the relevant Boltzmann equation for energy transfer from the SM to the dark sector is
\begin{align}
	\int \frac{d^3p_\varphi}{(2\pi)^3} E_\varphi \frac{df_\varphi}{dt} + \int \frac{d^3p_\alpha}{(2\pi)^3} p_\alpha \frac{df_\alpha}{dt} + \int \frac{d^3p_N}{(2\pi)^3} E_N \frac{df_N}{dt} &= \int \frac{d^3p_\phi}{(2\pi)^3} E_\phi \sum_i S_{\phi,i} \mathcal{C}_{\phi,i}[\text{SM} \to \phi] \,,
	\label{boltzScalar}
\end{align}
where $\sum_i S_{\phi,i} \mathcal{C}_{\phi,i}[\text{SM} \to \phi]$ refers schematically to the sum of all collision terms for the production of the $\phi$ sector and $S_{\phi,i}$ is the number of $\phi$ produced in the given process (note $S_{\phi , i}$ is negative if $\phi$ particles are destroyed). 
The collision term for a final 
state particle in a $2 \to 2$ process, $\mathcal{C}_a[ij\to ab]$, is 
\begin{equation}
    \mathcal{C}_a = \frac{1}{S } \frac{1}{2E_a} \int d\Pi_b \int d \Pi_i \int d\Pi_j \, (2\pi)^4 \delta^{(4)} (p_i + p_j - p_a - p_b) |\mathcal{M}(ij \to ab)|^2 f_i f_j (1\pm f_a)(1\pm f_b)  \, ,
    \label{collision}
\end{equation}
where $S$ is the symmetry factor which accounts for the multiplicities in the initial and final states, and $d\Pi_X = g_X d^3 p_X/((2\pi)^3 2E_X)$ denotes the integration over the phase space of particle $X$ with $g_X$ internal degrees of freedom. 
$|\mathcal{M}(ij\to ab)|^2$ is the squared matrix element averaged over initial and final state spins. The reverse process is written separately and enters the Boltzmann equation with a minus sign because particles are destroyed. The index of $\mathcal{C}$ denotes the particle which is singled out in the Boltzmann equation. 
The assumption of kinetic equilibrium implies that the $\varphi\leftrightarrow \alpha$, $\varphi \leftrightarrow N$ and $\alpha\leftrightarrow N$ collision terms that should appear in the right-hand side of Eqn.~\eqref{boltzScalar} in fact cancel out, for instance 
\begin{align}
\int \frac{d^3p_\varphi}{(2\pi)^3} E_\varphi \mathcal{C}_{\varphi}[\varphi \to \alpha \alpha] &= 2 \int \frac{d^3p_\alpha}{(2\pi)^3} p_\alpha \mathcal{C}_{\alpha}[\alpha \alpha \to \varphi]  \\
2 \int \frac{d^3p_\alpha}{(2\pi)^3} p_\alpha \mathcal{C}_{\alpha}[\alpha \alpha \to NN] &= 2 \int \frac{d^3p_N}{(2\pi)^3} E_N \mathcal{C}_{N}[NN \to \alpha \alpha]  ~.
\end{align}
The sterile neutrino number density after it thermalises is
\begin{equation}
n_N = \frac{3 \zeta(3) T_\phi^3}{2 \pi^2} ~,  
\end{equation}
since it has a Fermi-Dirac (FD) distribution with temperature $T_\phi$. 
The dark sector interactions will inevitably go out of equilibrium when $T_\phi \gg m_N$, thus the sterile neutrino maintains a relativistic number density.
Its relic energy density is therefore given by
\begin{equation}
    \Omega_N h^2 \simeq 0.3 \left( \frac{z}{0.1} \right)^3 \frac{m_N}{10 \text{ keV}} ~.
    \label{relicabundance}
\end{equation}
The evolution of $z$ with time will be computed in the following section. 
From Eqn.~\eqref{relicabundance}, we see that for a keV-scale sterile neutrino, $z \simeq 0.1$ at very late times generates the observed DM abundance, $\Omega_{\rm dm} h^2 = 0.12$ \cite{Planck:2018vyg}. 
We will find in Sec.~\ref{CalcTphi} that $z \propto \sqrt{\kappa}$ and is independent of $m_N$. 
Thus, the relation $\Omega_N h^2 \propto \kappa^{3/2} m_N$ removes one free parameter from the model if we insist on a relic sterile neutrino density in agreement with observations.

Strictly, $T_\phi$ is ill-defined after the $\phi-N$ bath goes out of equilibrium, and consequently $z$ is also ill-defined. 
However, after decoupling the $\alpha$ and $N$ free-stream with $p \propto a^{-1}$, so after the decoupling time, $t_{\phi , \text{dec}}$, we set $T_\phi(t) \equiv [a(t_{\phi ,\text{dec}})/a(t)] T_\phi(t_{\phi ,\text{dec}})$, 
In this case, the evolution of $z$ after $t_{\phi,\text{dec}}$ is determined only by the number of relativistic SM degrees of freedom in entropy, $g_*^S$, and becomes constant after electron decoupling.

Finally, we note that in principle the pNGB $\alpha$ also contributes to the DM abundance due to its tiny mass, however this abundance is suppressed by a factor $\sim m_\alpha/m_N \ll 1$ compared to the sterile neutrino abundance and is therefore completely negligible.

\subsection{Contribution to \texorpdfstring{$N_\text{eff}$}{Neff}}
There is a contribution to $N_\text{eff}$ from both $\alpha$ and $N$, which is easily calculable in this scenario. 
Their energy density at $T_{\phi} \gg m_N$ is
\begin{equation}
    \rho_\mathrm{dark} = g_*^\mathrm{dark} \frac{\pi^2 T_\phi^4}{30} ~,
    \label{rhodark}
\end{equation}
where $g_*^\mathrm{dark} =  11/4$. 
The value of $N_\text{eff}$ at $T = 1$ MeV is therefore
\begin{equation}
    N_\text{eff} = \rho_\mathrm{dark} \left( \frac{7}{8} (4/11)^{4/3} \rho_\gamma \right)^{-1} \simeq 12.1 \, \left[z(1 \text{ MeV})\right]^4 ~.
\end{equation}
Taking $z \lesssim 0.1$, as required for the correct relic abundance (cf.~Eqn.~\eqref{relicabundance}), leads to $N_\text{eff} \sim 10^{-3}$, too small to be detected even in the next generation of experiments \cite{CMB-S4:2016ple}, which have an expected sensitivity of $\sigma(N_{\rm eff})\simeq 0.02-0.03$.

\subsection{Sterile neutrino free-streaming horizon}

The average momentum of the sterile neutrinos, given its FD distribution, is
\begin{align}
    \left \langle p_N (T_\phi) \right\rangle &= \frac{\int d^3 p f_N p}{\int d^3 p f_N } 
    = 3.15 T_{\phi} ~.
\end{align}
We can estimate the temperature at which the sterile neutrinos become non-relativistic by $\left \langle p_N  \right\rangle = m_N$, i.e. $T_{\phi,\mathrm{nr}} = m_N/3.15$. 
We denote the corresponding time by $t_\mathrm{nr}$. 
Then the free-streaming horizon is given by 
\begin{equation}
    \lambda_{FS} = \frac{\sqrt{t_{\mathrm{nr}} t_{\mathrm{eq}}}}{a_\mathrm{eq}} \left( 5 + \ln \frac{t_\mathrm{eq}}{t_\mathrm{nr}} \right) 
    \simeq 366 \frac{z\, \mathrm{keV}}{m_N} \left(7.299 + \ln \frac{m_N}{z\, \mathrm{keV}}\right) \mathrm{kpc}~,
    \label{lambdafs}
\end{equation}
following the calculation of Ref.~\cite{Adulpravitchai:2014xna}, where $t_{\mathrm{eq}} = 1.9 \times 10^{11}$s is the time at matter-radiation equality and $a_\text{eq} = 8.3 \times 10^{-5}$ is the corresponding scale factor. 
Bounds from Ly-$\alpha$ data set a lower limit of $m_{\rm DM} \geq 5.3$ keV on early-decoupled fermionic DM \cite{Irsic:2017ixq}, which can be converted into the constraint $\lambda_{FS} \lesssim 66$ kpc. 
Indeed, the bound on the sterile neutrino mass is also $m_N \gtrsim 5.3$ keV, because for a given value of $m_N$, producing the observed dark matter abundance fixes the temperature ratio $z$ via Eqn. \eqref{relicabundance}, and thus the free-streaming horizon. 
This lower limit is shown in Fig.~\ref{fig:Nthermalisation}.

\section{Heating of the \texorpdfstring{$\phi$}{phi} bath: Calculation of \texorpdfstring{$T_\phi$}{Tphi}}
\label{CalcTphi}
In the previous section, we introduced the time-dependent temperature ratio, $z$, and showed that it should be $\mathcal{O}(0.1)$ for the correct sterile neutrino relic abundance. 
In this section, we compute this function explicitly by calculating the energy injection into the $\phi-N$ thermal bath. 
The integrated Boltzmann equation is
\begin{equation}
    \frac{d\rho_\mathrm{dark}}{dt} + 4H \rho_\mathrm{dark} = \int 
    \frac{d^3 p_\varphi}{(2\pi)^3} E_\varphi
\sum_i S_{\varphi,i} \mathcal{C}_{\varphi,i}[\text{SM} \to \varphi] \, ,
\label{integratedboltzT}
\end{equation}
where $i$ sums over all channels of $\varphi$ production $\text{SM} \to \varphi$ ($\alpha$ production is subdominant, and recall that $N$ are created from the $\phi$ sector). 
The energy density of the thermalised dark sector is given in Eqn.~\eqref{rhodark}, where $g_*^\text{dark} = 15/4$ for $T_\phi \gg m_\varphi$ and $11/4$ for $T_\phi \ll m_\varphi$. 
Hence the left-hand side of Eqn.~\eqref{integratedboltzT} is
\begin{align}
    \frac{d\rho_\mathrm{dark}}{dt} + 4H \rho_\mathrm{dark} 
    &= \frac{2 \pi^2 g^*_\mathrm{dark} m_h^6 H_* z^3}{15 x^5} \frac{dz}{dx} \, ,
    \label{boltzLHS}
\end{align}
where $H_* \equiv H/T^2$ and we parameterise time in terms of
\begin{equation}
    x\equiv \frac{m_h}{T}\;.
\end{equation}
Different channels of $\varphi$ production are only kinematically allowed for certain values of $m_\varphi$. 
Combining Eqns.~\eqref{integratedboltzT} and \eqref{boltzLHS}, we obtain
\begin{align}
    z(x) = \left[ \int_0^x dx' \frac{30 x'^5}{\pi^2 g^*_\mathrm{dark} m_h^6 H_*} \left( \sum \limits_i \int \frac{d^3 p_\varphi}{(2\pi)^3} E_\varphi S_{\varphi,i} \mathcal{C}_{\varphi,i}[\text{SM} \to \varphi] \right) \right]^{1/4} \, ,
    \label{zeq}
\end{align}
where we sum over the allowed processes and $S_{\varphi,i}$ denotes the number of $\varphi$ particles produced in process $i$. 
We now compute these collision terms.

\subsection{Higgs decays}

For $m_\varphi < m_h/2$, Higgs decays dominate $\varphi$ production. 
The integrated collision term is
\begin{align}
2 \int \frac{d^3p_\varphi}{(2\pi)^3} E_\varphi \mathcal{C}_{\varphi}[h \to \varphi \varphi] &= m_h n_h^{eq} \Gamma(h \to \varphi \varphi) = \frac{\kappa^2 m_h^5 K_2(x)}{128 \pi^3 \lambda x} \left( \frac{m_h^2 + 2m_\varphi^2}{m_h^2 - m_\varphi^2} \right)^2 \sqrt{1 - \frac{4m_\varphi^2}{m_h^2}} \, , 
\label{zRHSa} 
\end{align}
with
\begin{equation}
    \Gamma(h\to \varphi\varphi) = \frac{\kappa^2 v^2}{32\pi m_h} \left(\frac{2m_\varphi^2 + m_h^2}{m_h^2-m_\varphi^2}\right)^2\sqrt{1-\frac{4m_\varphi^2}{m_h^2}} \, .
\end{equation}
In the limit that this decay is the only relevant process, combining Eqns.~\eqref{zeq} and \eqref{zRHSa}, gives
\begin{align}
z(x) &\simeq 
0.018 \sqrt{ \frac{\kappa}{10^{-10}} \frac{ 2m_\varphi^2 + m_h^2}{m_h^2 - m_\varphi^2} } \left( \frac{g_*^s}{100}\right)^{1/3} \left( \frac{100}{g_*^\rho} \right)^{1/8} \left( \frac{3.75}{g^*_\mathrm{dark}} \right)^{1/4}  \left(1 - \frac{4m_\varphi^2}{m_h^2} \right)^{1/8} \notag \\&\times
\left( \int_{x_{\rm ew,c}}^x dx' x'^4 K_2(x') \right)^{1/4} \, ,
\label{zLight}
\end{align}
where the integral's lower limit at $x_{\rm ew,c}\equiv m_h/T_{\rm ew,c}=0.78$ reflects the electroweak phase transition, after which $h \to \varphi \varphi$ is allowed. 
The value of the integral at late times is $\int_{0.78}^\infty dx' x'^4 K_2(x') = 23.3$.  
Note that entropy dilution of the SM thermal plasma has explicitly been taken into account. We use the analytic expressions in the Appendix of \cite{Wantz:2009it} to trace the evolution of $g_*^s$ and $g_*^\rho$ accurately for our numerical results.

\subsection{\texorpdfstring{$\text{SM SM} \to \varphi \varphi$}{SM SM -> varphi varphi} scattering}
\label{ssec:smscattering}

For all values of $m_\varphi$, SM particles can scatter to produce pairs of $\varphi$. 
Before the electroweak phase transition, the cross section for the scattering of a Higgs doublet with its antiparticle to a particle-antiparticle pair of dark sector particles, $H H^* \to \phi \phi^*$, is described by the cross section 
\begin{equation}
    \sigma(HH^* \to \phi\phi^*) = \frac14 \frac{\kappa^2}{16\pi s} \, ,
\end{equation}
neglecting all masses. After the electroweak phase transition, there are three such annihilations at tree-level: $hh, VV, f\bar{f} \to \varphi \varphi$, where $V$ is a vector gauge boson that could be either the $W$ or $Z$, and $f$ is any SM fermion. 
Both $VV$ and $f\bar{f}$ scattering are $s$-channel processes mediated by the Higgs, consequently $t \bar{t}$ scattering is much larger than any other fermion scattering due to the dominance of $y_t$ over all other Yukawas. 
The production of $\varphi$ from these annihilations is subdominant compared to the three-body processes $h \to \varphi \varphi$ and $hh \to \varphi$ when either is kinematically allowed. 
However, for $m_h/2 \leq m_\varphi \leq 2m_h$ only scattering is allowed, therefore it entirely determines the injection of energy into the $\phi$ bath.

\subsubsection*{Higgs scattering}

There are seven tree-level diagrams which mediate $hh \to \varphi \varphi$. 
The three dominant diagrams are the contact interaction and the two s-channel diagrams, one with a virtual $h$, the other with a virtual $\varphi$. 
The two t-channel diagrams and two u-channel diagrams (with virtual $h$ or $\varphi$) are suppressed by an extra factor of $\kappa \ll 1$ and therefore negligible. 
The squared matrix element is then
\begin{align}
    \left|\mathcal{M}(hh \to \varphi \varphi)\right|^2 &\simeq \frac{\kappa^2 (s+2m_h^2)^2 (s+2m_\varphi^2)^2}{(s-m_h^2)^2 (s-m_\varphi^2)^2},
    \label{hhppMat}
    \end{align}
    where we dropped terms involving the $h$ and $\varphi$ widths, since $\Gamma_{h,\varphi} \ll m_{h, \varphi}$ and $s \geq \max\{4m_h^2, 4m_\varphi^2\}$. 
    The collision term is
\begin{align}
      2 \int \frac{d^3p_\varphi}{(2\pi)^3} E_\varphi \mathcal{C}_{\varphi}[hh \to \varphi \varphi] 
    &= \frac{m_h^5}{2^{11} \pi^5 x^5} \int_{4x^2}^\infty dr \sqrt{ r - 4 x_{\varphi T}^2} \sqrt{r - 4 x^2} K_1(\sqrt{r}) |\mathcal{M}(hh \to \varphi \varphi)|^2 \Theta(r - 4x_{\phi T}^2) ~, 
    \label{zRHSb1}
\end{align}
where $x_{\varphi T} = m_\varphi/T$, taking a Maxwell-Boltzmann (MB) distribution for the Higgs. 
Using a Bose-Einstein (BE) distribution instead gives a similar result.

\subsubsection*{Vector boson scattering}
Vector-boson scattering is an $s$-channel processes mediated by the Higgs, with cross-section
\begin{equation}
    \sigma(VV \to \varphi \varphi) = \frac{1}{9} \frac{\kappa^2}{32 \pi s} \sqrt{\frac{s - 4m_\varphi^2}{s - 4 m_V^2}} \frac{s^2 - 4s m_V^2 + 12 m_V^4}{(s-m_h^2)^2 } \, ,
    \label{sigmaVVphiphi}
\end{equation}
averaging over polarisations, where we again neglect the Higgs decay width.  Then the collision term is
\begin{align}
2 \int \frac{d^3p_\varphi}{(2\pi)^3} E_\varphi \mathcal{C}_{\varphi}[VV \to \varphi \varphi] &= \frac{ \kappa^2 m_h^5}{2^{11} \pi^5 x^5} \sum \limits_V \int_{4 x_V^2}^\infty dr \sqrt{r - 4x_V^2} \sqrt{r - 4x_{\varphi T}^2} 
\frac{12 x_V^4 - 4 r x_V^2 + r^2}{(r - x^2)^2 }
\notag
\\&\times
K_1(\sqrt{r}) \Theta(r - 4x_{\varphi T}^2)  \, , \label{zRHSb2}
\end{align}
where $x_V = m_V/T$, using a MB distribution for the vector bosons.

\subsubsection*{Top quark scattering}
Like vector-boson scattering, $t\bar{t}$ scattering occurs via an $s$-channel diagram with a virtual Higgs. 
The cross-section is
\begin{align}
    \sigma(t \overline{t} \to \varphi \varphi) 
    = \frac{1}{192 \pi s} \frac{ \kappa^2 m_t^2 \sqrt{(s - 4m_\varphi^2)(s - 4m_t^2)}}{(s - m_h^2)^2 } ~,
    \label{sigmattphiphi}
\end{align}
averaging over initial spins and colours. 
The collision term is
\begin{align}
    2 \int \frac{d^3p_\varphi}{(2\pi)^3} E_\varphi \mathcal{C}_{\varphi}[t \overline{t} \to \varphi \varphi] 
    &
    = \frac{6 \kappa^2 m_t^2 m_h^3}{ 2^{10} \pi^5 x^3} \int_{4x_t^2}^\infty dr \frac{\sqrt{r - 4x_{\varphi T}^2} (r - 4x_t^2)^{3/2}}{(r - x^2)^2} K_1(\sqrt{r}) \Theta(r - 4 x_{\varphi T}^2)  \, ,
    \label{zRHSb3}
\end{align}
where $x_t = m_t/T$ and $\tilde{\Gamma}_h = \Gamma_h/T$, taking a MB distribution for the top.

The collision terms from Higgs, vector boson and top quark scattering are all approximately the same order of magnitude since $\lambda \sim g_{1,2} \sim y_t$, therefore no process can be ignored. 
As mentioned above, scatterings involving lighter fermions are suppressed by their small Yukawa couplings, $y_f \ll y_t$, and are therefore negligible.

\subsection{\texorpdfstring{$hh \to \varphi$}{h h -> varphi} inverse decay}
When $m_\varphi > 2m_h$ and $T_\phi<T_{\phi,c}$, the $hh \to \varphi$ process opens up and in fact dominates the annihilations considered just above because these 4-body processes are relatively suppressed by an additional phase space factor. 
Since $\varphi$ production is peaked at around $T \sim m_\varphi$, at which time the Higgs may be relativistic, we take a BE distribution for the Higgs here. 
Then we obtain the collision term, 
\begin{align}
    \int \frac{d^3p_\varphi}{(2\pi)^3} E_\varphi \mathcal{C}_{\varphi}[hh \to \varphi] 
    &= \frac{m_h^3 |\mathcal{M}( hh \to \varphi)|^2}{32 \pi^3 x^3} \int^\infty_{x_{\varphi T}} dF_{\varphi T} \frac{F_{\varphi T}}{e^{F_{\varphi T}}-1} \ln \Bigg[ \mathrm{csch} \left( \frac{F_{\varphi T} - y_{\varphi T} \sqrt{1 - \frac{4m_h^2}{m_\varphi^2}}}{4} \right) \notag \\
    &\times \sinh\left( \frac{F_{\varphi T} + y_{\varphi T} \sqrt{1 - \frac{4m_h^2}{m_\varphi^2}}}{4} \right) \Bigg] \, ,
    \label{zRHSc}
\end{align}
where here $F_{\varphi T} = E_\varphi/T$ and $y_{\varphi T} = p_\varphi/T$, with the squared matrix element
\begin{equation}\label{eq:Mhhphi}
   | \mathcal{M}(hh \to \varphi)|^2 = \kappa^2 v_\phi^2 \left(\frac{2m_h^2 + m_\varphi^2}{m_\varphi^2 - m_h^2} \right)^2\, .
\end{equation}

\subsection{Remarks on the dark sector temperature ratio \texorpdfstring{$z$}{z}}
Having computed all the collision terms, we can insert Eqns.~\eqref{zRHSa}, \eqref{zRHSb1}, \eqref{zRHSb2}, \eqref{zRHSb3}, and \eqref{zRHSc} into Eqn.~\eqref{zeq} to find $z$ in general. 
We can make several observations about the form of $z$. 
Firstly, since all collision terms are proportional to $\kappa^2$, we have from Eqn.~\eqref{zeq} that $z \propto \sqrt{\kappa}$. 
Moreover, the collision terms are all independent of $m_N$, therefore so is $z$. 
This was already mentioned in Section \ref{sec:Nproduction}, where we argued that the relic abundance scales as $\Omega_N h^2 \propto \kappa^{3/2} m_N$.

Secondly, the collision terms are positive and approach 0 as $x$ grows much larger than 1. 
This means that $z$ increases with $x$, up to shifts in $g_*$ due to entropy dilution of the SM thermal bath as SM fields freeze out. 
However, at some $m_N \ll T \ll m_h$, the energy injection into the $\phi-N$ bath becomes tiny and consequently $z$ reaches its maximum, after which it decreases slightly due to this entropy dilution. 

\begin{figure}
    \centering
    \includegraphics[width=0.8\textwidth]{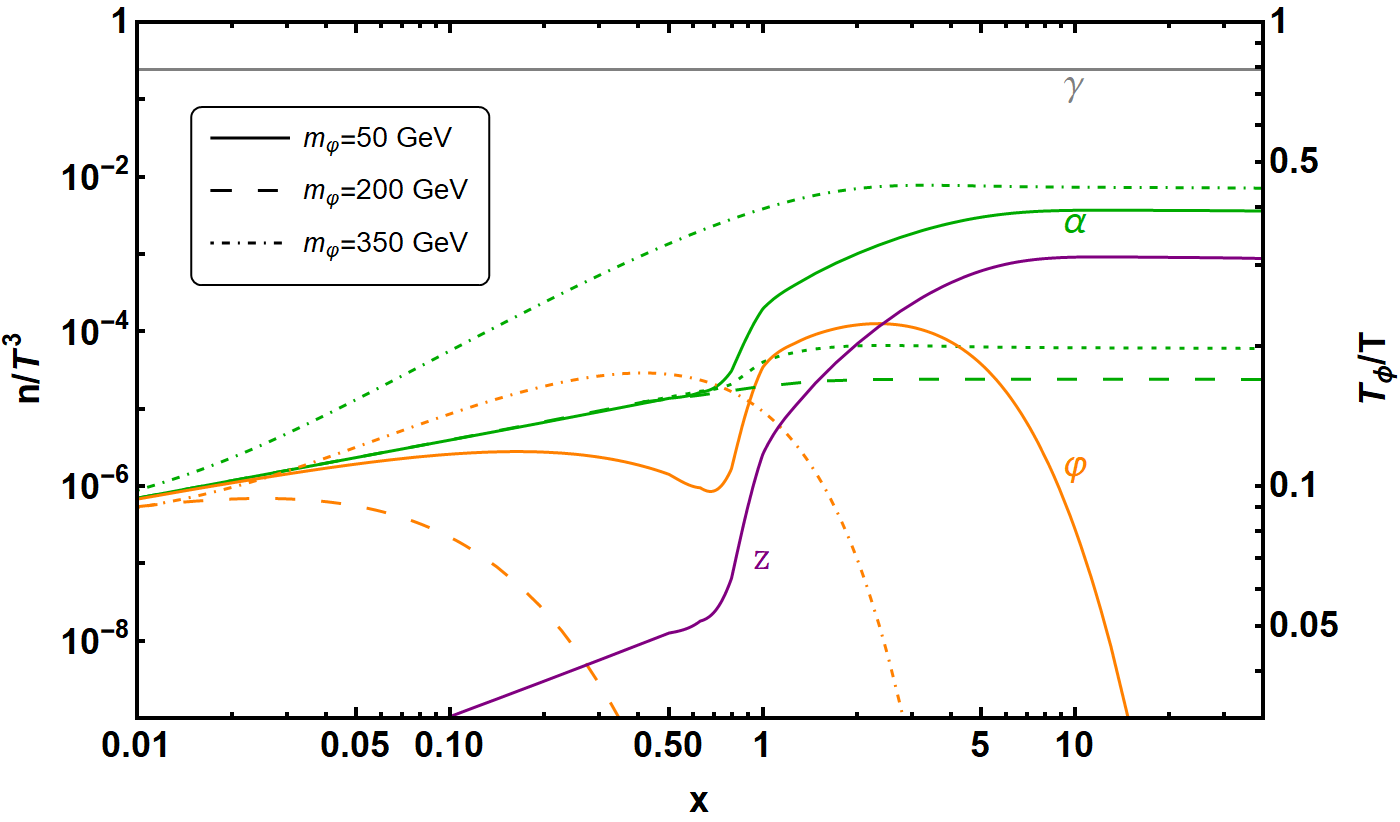}
    \caption{Number densities of $\alpha$ (green) and $\varphi$ (orange) for different values of $m_\varphi=50(200)[350]$ GeV using solid (dashed) [dotdashed] lines, with $\lambda_\phi = 0.1$ and $\kappa = 10^{-8}$. We show the evolution of $z$ for $m_\varphi=50$ GeV in purple. For the $m_\varphi = 50$ and 350 GeV cases, the $\phi$ sector remains in equilibrium until production is complete, i.e. until $n_\alpha/T^3$ reach their final values. 
    For the $m_\varphi = 200$ GeV case, equilibration only lasts until $x \simeq 0.4$. 
    The dashed green line shows the true $\alpha$ number density while the dotted green line gives its value had thermalisation persisted. 
    }
    \label{numberplot}
\end{figure}

In Fig.~\ref{numberplot} we illustrate the evolution of the number densities of the two new scalars, $\alpha$ (green) and $\varphi$ (orange), for fixed couplings in the scalar potential, $\lambda_\phi=0.1$ and $\kappa=10^{-8}$. 
We represent different scalar masses, $m_\varphi = 50 (200) [350]$ GeV, by the solid (dashed) [dotdashed] lines. For all three masses, the number density of the pNGB $\alpha$ normalised to $T^3$ approaches a constant value for large $x$. As $\alpha$ is in thermal equilibrium, its number density is fixed by the temperature ratio $z$, with $n_\alpha \sim z^3$. 
Note that $n_\alpha$, and hence $z$, is larger for $m_\varphi = 50, 350$ GeV than for the intermediate mass, $m_\varphi = 200$ GeV. This is due to the efficient $h \to \varphi \varphi$ decay (for $m_\varphi\leq m_h/2$) and $hh \to \varphi$ inverse decay ($m_\varphi\geq 2m_h$) processes, respectively. 
All number densities initially increase relative to the photon number density. The $\alpha$ number density freezes in for $x\sim 1-5$ and the $\varphi$ number density is exponentially suppressed when $\varphi$ becomes non-relativistic at $m_\varphi \simeq T_\phi = z m_h/x$, i.e. at $x\simeq z m_h/m_\varphi$. 
Note that the exponential suppression when $m_\varphi=350$ GeV occurs later than when $m_\varphi=200$ GeV\textemdash this is due to the larger temperature ratio $z$ in the former case. 

For $m_\varphi = 50$ and $350$ GeV, the $\phi$ sector is in equilibrium throughout the period of its production and therefore the $\varphi$ and $\alpha$ are well described by thermal number densities. 
On the other hand, for $m_\varphi = 200$ GeV, the $\varphi \varphi \leftrightarrow \alpha \alpha$ rate drops out of equilibrium around $x \simeq 0.4$ and a non-zero chemical potential develops for $x\gtrsim 0.4$, while kinetic equilibrium is maintained. Consequently, the $\alpha$ number density is approximately a factor of $2.5$ smaller than it would be had thermalisation endured throughout its production, as shown by the discrepancy between the dashed green line (the true number density) and the dotted one (number density assuming thermalisation). 

Having computed the evolution of $T_\phi$, we now investigate the conditions under which dark sector thermalisation occurs.

\section{Thermalisation of the \texorpdfstring{$\phi$}{phi} bath}
\label{sec:thermalisationPhi}
In this section we consider the thermalisation of the $\phi$ bath. 
The assumption of thermalisation was used to compute the sterile neutrino yield analytically, to calculate the temperature evolution of the bath encoded in $z(x)$, and to find $N_{\mathrm{eff}}$ and $\lambda_{FS}$. 
This assumption must of course be verified. 
The strength of $\phi$ sector interactions is characterised by the quartic coupling, $\lambda_\phi$.

For a process $a+X \leftrightarrow Y$ with X,Y denoting an arbitrary number of particles, the thermally-averaged collision rates which are relevant for chemical equilibrium of particle species $a$ are defined as 
\begin{align}
    \langle \Gamma(a X \leftrightarrow Y) \rangle_a
    &= \frac{1}{n_a}\int \frac{g_a d^3 p_a}{(2\pi)^3} \mathcal{C}_a [a X \leftrightarrow Y]  \, ,
\end{align}
where $\mathcal{C}_a$ denotes the collision term introduced in Eqn.~\eqref{collision}. 
A similar expression applies for the inverse process.
For kinetic equilibrium, we are concerned with the relaxation rate $\Gamma_{\rm relax}$. 
For a relativistic particle, it is given by the collision rate, $\Gamma_{\rm coll}=\langle \Gamma \rangle$, while for a non-relativistic particle it receives a suppression and is given by \cite{Hofmann:2001bi}
\begin{equation}
 \Gamma_\text{relax,a} \simeq (T/m_a) \, \Gamma_{\rm coll,a} \, .
\end{equation}

For $T_{\phi}>T_{\phi,c}$, the relevant processes for chemical and kinetic equilibration are $\phi\phi \to \phi\phi\phi\phi^*$ and $\phi\phi\to \phi\phi$, together with processes where some $\phi$ are replaced by the antiparticle $\phi^*$. We use the results in \cite{Arcadi:2019oxh,DeRomeri:2020wng} to estimate thermalisation for the intermediate ranges above the critical temperature. 
More importantly, for $T_\phi<T_{\phi,c}$ the relevant processes are $\varphi\leftrightarrow \alpha\alpha$, $\varphi\varphi\leftrightarrow\alpha\alpha$, and $\alpha \varphi\to\alpha\varphi$, whose rates are
\begin{align}
    \Gamma(\varphi\to\alpha\alpha) & = \frac{\lambda_\phi m_\varphi}{16\pi}  \\
    \sigma(\varphi\varphi \to\alpha\alpha) &= \frac{\lambda_\phi^2 }{4\pi \sqrt{s(s-4m_\varphi^2)}}
    \\\nonumber & \times
    \left[1 + 
          \frac{(2m_\varphi^2+s)^2}{2(m_\varphi^2-s)^2} 
         + \frac{3m_\varphi^4+m_\varphi^2 s-s^2}{s^2-3m_\varphi^2 s +2 m_\varphi^4} \frac{4 m_\varphi^2}{\sqrt{s(s-4m_\varphi^2)}}
        \mathrm{arctanh}\left(\frac{2\sqrt{s(s-4m_\varphi^2)}}{2s-4 m_\varphi^2}\right)  \right]
    \\
    \sigma(\alpha\alpha\to\varphi\varphi) & =\sigma(\varphi\varphi\to\alpha\alpha) \left(1-\tfrac{4m_\varphi^2}{s}\right) 
    \\\label{eq:sigma_aPaP}
    \sigma(\alpha\varphi\to\alpha\varphi)
    &\simeq
    \frac{\lambda_\phi^2}{4\pi m_\varphi^2}
    \Bigg( \frac{s^6+4m_\varphi^2s^5+29 m_\varphi^4s^4+4m_\varphi^6s^3+29 m_\varphi^8s^2 +4 m_\varphi^{10}s+m_\varphi^{12}}{3 m_\varphi^4 s^2( s^2-m_\varphi^2 s+m_\varphi^4)} \nonumber
    \\&\qquad\qquad\qquad
    -\frac{6( s+m_\varphi^2)^2}{( s-m_\varphi^2)^2} \ln\frac{ s^2-m_\varphi^2 s+m_\varphi^4}{m_\varphi^2 s}\Bigg) \, .
   \end{align}
We neglected the $u$-channel contribution in the cross section for $\alpha \varphi \to \alpha\varphi$, where the self-energy plays an important role. 
These issues are discussed in App.~\ref{app:selfenergy}, where we give the expressions for this self-energy, and App.~\ref{app:aPaP}, where we explain that this contribution is generally negligible but is very computationally taxing.
For $\varphi \to \alpha \alpha$ and $\varphi \varphi \to \alpha \alpha$, the relaxation rate is equal to the collision rate $\Gamma_{\mathrm{relax}} = \Gamma_\text{coll}$, while for processes with the scalar $\varphi$ in the final state, it is given by $\Gamma_{\mathrm{relax}} \approx \Gamma_{\rm coll}/x_\varphi$, where 
\begin{equation}
x_\varphi \equiv \frac{m_\varphi}{T_\phi}\;.
\end{equation}
When all relevant interaction rates are faster than the Hubble rate, thermalisation is guaranteed. Hence, the 
conditions for thermalisation are
\begin{align}
    \langle \Gamma(\varphi \leftrightarrow \alpha\alpha)\rangle_\alpha &> H   \label{phitherm1} \\
    \langle \Gamma(\varphi\varphi \leftrightarrow \alpha\alpha)\rangle_{\alpha} &> H  \label{phitherm2}
    \\
    x_\varphi^{-1} \langle \Gamma(\alpha\varphi\to\alpha\varphi)\rangle_\varphi &> H\, ,  \label{phitherm3}
\end{align}
where the conditions in the first two lines ensure kinetic and chemical equilibrium for $\alpha$ and chemical equilibrium for $\varphi$. The corresponding interaction rates in the Boltzmann equations of $\varphi$ are automatically satisfied in this case. The last condition ensures kinetic equilibrium of $\varphi$. 
We will demonstrate the thermalisation in two stages. First we find when a non-thermal distribution described by a delta function thermalises quickly enough, then we calculate when a thermal distribution of $\varphi$ and $\alpha$ particles remains in thermal equilibrium. 

Note that the pNGB plays an important role in thermalising the dark sector, in fact takes part in all relevant processes. This is the main difference to the thermalisation of the real scalar field. In \cite{DeRomeri:2020wng} the authors find that for the heavy scalar, analogous to our $\varphi$, thermalisation can be achieved via $\varphi\varphi \to \varphi\varphi\varphi\varphi\varphi$ but only for larger couplings, e.g.~for $m_\varphi\gtrsim 100$ GeV they require $\kappa \gtrsim 10^{-9}$ and $\lambda_\phi\gtrsim 0.1$.

\subsection{Initial \texorpdfstring{$\phi$}{phi} thermalisation}
The thermal averaging depends on the distribution functions of the particles involved in the given interaction. 
If they are already in equilibrium, we can take a MB, BE or FD distribution as appropriate. 
If the particles are not in equilibrium, however, their distribution functions can in general only be found by numerically solving a Boltzmann equation. 
To avoid this complication, we will assume below that at early times the $\varphi$, $\alpha$, and $N$ have delta function distributions. This is motivated by the instantaneous decay approximation. For a homogeneous distribution of non-relativistic Higgs bosons, the distribution of $\varphi$ in the instantaneous decay approximation is homogeneous and isotropic with a definite energy, since $h$ is a scalar which dominantly decays at rest. 
After averaging over the homogeneous and isotropic initial states, the resulting distributions for inverse decays and $2\to 2$ scattering processes are also homogeneous and isotropic. 
In order to obtain semi-analytic estimates, and to model an extremely non-thermal distribution, we assume that the $\varphi$ distributions for inverse decay and $2\to 2$ scattering processes have a definite energy which is determined by the centre of mass energy of the collisions. 
This neglects the velocity dispersion of the initial state particles, which leads to a broadening of the $\varphi$ distribution. 
A detailed discussion requires to solve the Boltzmann equations numerically and is beyond the scope of this study. 
Consider non-thermal distributions described by the delta functions,
\begin{align}
    f_\alpha(p_\alpha) &= C_\alpha \delta(p_\alpha - X_\alpha) &
    f_\varphi(E_\varphi) & = C_\varphi \delta(E_\varphi - X_\varphi)  \, ,
\end{align}
with $p_\alpha\equiv |\vec p_\alpha|$. 
The number densities are
\begin{align}
    n_\alpha &= \int\frac{d^3 p_\alpha}{(2\pi)^3} f_\alpha = \frac{C_\alpha X_\alpha^2}{2\pi^2} 
    &
    n_\varphi &= \int\frac{d^3 p_\varphi}{(2\pi)^3} f_\varphi = \frac{C_\varphi X_\varphi\sqrt{X_\varphi^2-m_\varphi^2}}{2\pi^2} ~,
    \label{cacphi}
\end{align}
and thus $C_\alpha=2\pi^2 n_\alpha/X_\alpha^2$ and $C_\varphi =2\pi^2 n_\varphi/\left(X_\varphi^2 \sqrt{1-m_\varphi^2/X_\varphi^2}\right)$. 
With the $\varphi$ and $\alpha$ distribution functions thus defined, the thermally-averaged interaction rates are 
\begin{align}
    \langle\Gamma(\varphi \to \alpha\alpha) \rangle_{\alpha} &= \frac{C_\varphi m_\varphi X_\varphi}{ \pi^2 n_\alpha} \Gamma(\varphi\to \alpha\alpha) \sqrt{1-\frac{m_\varphi^2}{X_\varphi^2}} \\ \langle\Gamma(\alpha\alpha\to \varphi) \rangle_{\alpha} &= \frac{C_\alpha^2 m_\varphi}{ {2} \pi^2 n_\alpha} \Gamma(\varphi\to\alpha\alpha) \theta(2X_\alpha-m_\varphi)
\\
\langle\Gamma(\varphi\varphi\to \alpha\alpha)\rangle_{\alpha} &=  \frac{C_\varphi^2 m_\varphi^4}{ 4 \pi^4 n_\alpha} \int_{1}^{\frac{X_\varphi^2}{m_\varphi^2}} d\xi\,  \sigma\left(s=4m_\varphi^2\xi\right) \sqrt{\xi (\xi-1)} \\
\langle\Gamma(\alpha\alpha\to\varphi\varphi)\rangle_{\alpha} &= \frac{C_\alpha^2 m_\varphi^4}{ 4\pi^4 n_\alpha} \int_{1}^{\frac{X_\alpha^2}{m_\varphi^2}} d\xi\,  \sigma\left(s=4m_\varphi^2\xi\right) (\xi-1) \\
\langle\Gamma(\alpha\varphi\to\alpha\varphi)\rangle_\varphi &= 
\frac{C_\alpha C_\varphi m_\varphi^4}{32\pi^4 n_\varphi} \int_1^{\frac{(X_\varphi+X_\alpha)^2}{m_\varphi^2}} d\xi \, \sigma(s=m_\varphi^2 \xi)  (\xi - 1) 
\Theta\left( X_\varphi + X_\alpha - \frac{4X_\alpha^2 \xi + m_\varphi^2 (\xi - 1)^2 }{4X_\alpha (\xi - 1)} \right) \, .
\end{align}
Now, as an approximation, let us take $X_\varphi = \langle E_h \rangle/2$ when $m_\varphi < m_h/2$, $X_\varphi = \langle E_h \rangle$ when $2m_h \geq m_\varphi \geq m_h/2$, and $X_\varphi = 2 \langle E_h \rangle$ when $m_\varphi > m_h/2$, where $\langle E_h \rangle = \rho_h/n_h$. 
We make this $m_\varphi$-dependent approximation since, as discussed above, $\varphi$ production is dominated by $h \to \varphi \varphi$ in the smallest mass region, by $hh, t\bar{t}$ and $VV$ scattering to $\varphi \varphi$ in the intermediate regime, and by $hh \to \varphi$ in the heavy mass region. 
Then, since the $\varphi$ particles rapidly decay via $\varphi \to \alpha \alpha$, we take as the initial condition that the $\alpha$ energy is half the $\varphi$ energy, i.e. $X_\alpha = X_\varphi/2$. 
We note for $m_\varphi > m_h$ that at sufficiently late times $X_\varphi < m_\varphi$, which is unphysical. 
Therefore, we impose that equilibration must occur while $X_\varphi > m_\varphi$.

We compute the $\varphi$ number density by summing over the processes which produce it, outlined in Sec.~\ref{CalcTphi}, then assume that the $\phi$ sector interactions are sufficiently strong that the $\varphi$ and $\alpha$ number densities are approximately equal. Thus, given $n_{\varphi, \alpha}$ and $X_{\varphi, \alpha}$, we can find $C_{\varphi, \alpha}$ from Eqn.~\eqref{cacphi}. 
It enables a calculation of the various thermalisation rates listed above, thereby allowing us to find the region of parameter space in which the $\phi$ sector equilibrates, given these assumptions.

\subsection{Continued \texorpdfstring{$\phi$}{phi} thermalisation}
If the $\phi$ sector enters into equilibrium, we want to find for how long it maintains it. 
For this step, we can evaluate the collision rates given thermal distributions of $\varphi$ and $\alpha$.  
As the interaction rates with BE distributions are strictly larger than the ones with MB distributions, $f_{BE} > f_{MB}$, we will make a conservative estimate and approximate all distributions by MB ones, as well as neglecting the $(1+f)$ Bose-enhancement factors. 
Given this, we find
  \begin{align}
    \langle\Gamma(\varphi\leftrightarrow \alpha \alpha) \rangle_\alpha &= \frac{m_\varphi^3 K_1(x_\varphi)}{2\pi^2 x_\varphi\, n_\alpha} \Gamma(\varphi \to\alpha \alpha)  \\
     \langle\Gamma(\varphi\varphi\leftrightarrow\alpha\alpha)\rangle_\alpha &= \frac{m_\varphi^6}{2\pi^4\,x_\varphi\,n_\alpha} \int_{1}^\infty d\xi \sqrt{\xi}(\xi-1)\, K_1\left(2x_\varphi \sqrt{\xi}\right) \sigma(s=4 m_\varphi^2 \xi) \\
    \langle\Gamma(\alpha\varphi\to\alpha\varphi)\rangle_\varphi  &= \frac{ m_\varphi^6}{32\pi^4 x_\varphi n_\varphi} \int_1^\infty d\xi \, \frac{(\xi-1)^2}{\sqrt{\xi}} \sigma(s=m_\varphi^2 \xi) K_1(x_\varphi \sqrt{\xi}) \, . 
      \end{align}
Inserting these rates into Eqns.~\eqref{phitherm1}, \eqref{phitherm2} and \eqref{phitherm3} tells us for how long the $\phi$ sector remains in equilibrium. 
Note that $\langle \Gamma(\varphi  \to \alpha \alpha) \rangle_{\varphi} = \langle \Gamma(\alpha \alpha \to \varphi) \rangle_{\alpha}$ and $\langle \Gamma(\varphi \varphi \to \alpha \alpha) \rangle_{\varphi} = \langle \Gamma(\alpha \alpha \to \varphi \varphi) \rangle_{\alpha}$ by detailed balance. 

\subsection{Discussion}

\begin{figure}
    \centering
    \includegraphics[width=0.8\textwidth]{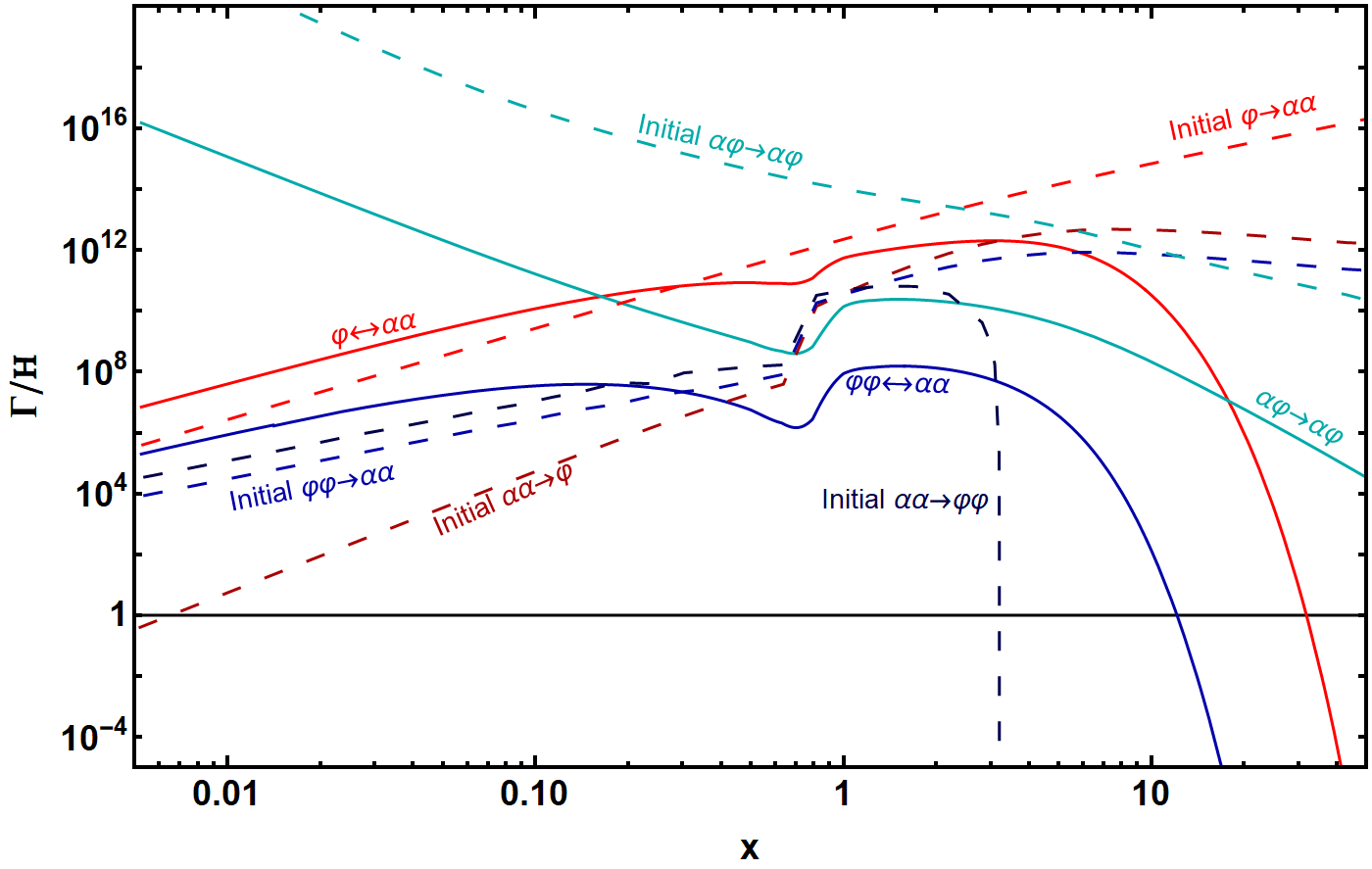}
    \caption{Relaxation rates in the $\phi$ sector normalised to the Hubble rate as a function of $x$ for $\lambda_\phi = 0.1, \kappa = 10^{-8}$, $m_\varphi = 50$ GeV. Continued (initial) thermalisation rates are shown as solid (dashed) lines. See the solid lines in Fig.~\ref{fig:evolplot} for the evolution of the different abundances. }
    \label{PhiRatesPlot}
\end{figure}

We illustrate the rates which are relevant for thermalisation for one particular benchmark point with $\lambda_\phi=0.1$, $\kappa=10^{-8}$ and $m_\varphi=50$ GeV in Fig.~\ref{PhiRatesPlot}. 
The figure displays the relaxation rates for $T_\phi<T_{\phi,c}$ for the different processes, normalised to the Hubble rate, as a function of $x$. 
The case for $T_\phi>T_{\phi,c}$ has been discussed in \cite{Arcadi:2019oxh} and the results presented in their Fig.~2. Taking into account the differences in the model (real vs complex scalar), we conclude that for $m_\phi=50$ GeV thermalisation is reached for $\lambda_\phi \gtrsim 10^{-3}$. 
This is consistent with Fig.~\ref{PhiRatesPlot}, which shows that thermalisation of the $\phi$ sector is easily achieved at early times.

In the following, we entirely focus on the rates for $T_\phi<T_{\phi,c}$ which determine the thermalisation of the $\phi$ sector at late times. 
The rates for initial thermalisation are shown as dashed lines and for continued thermalisation as solid ones. 
When all the dashed lines go above $\Gamma/H = 1$, there is initial equilibration: from this time on, the $\phi$ sector particles are well approximated by equilibrium distributions and the solid lines become the relevant ones.  
For $x<0.78$, electroweak symmetry is intact and the production of the $\phi$ sector particles only occurs via $H H^* \to \phi \phi^*$. 
Since the relaxation rates, $\Gamma$, are larger than the Hubble rate for $x<0.78$ in Fig.~\ref{PhiRatesPlot}, the dark sector quickly thermalises with a temperature $T_\phi =z\, T$. 
As $z\lesssim 0.1$, the scalar $\varphi$ becomes non-relativistic (cf. Fig.~\ref{numberplot}) and the interaction rates become Boltzmann suppressed, which is apparent from the drop in the continued relaxation rates around $x\lesssim 0.78$. 
After the electroweak phase transition at $x\geq 0.78$, Higgs decays become possible and indeed dominate the energy transfer to the dark sector, ultimately leading to increased relaxation rates since $T_\phi/T$ grows. 

The fast drop in the continued relaxation rates $\varphi\leftrightarrow \alpha\alpha$ and $\varphi\varphi\leftrightarrow\alpha\alpha$ originates from the Bessel function $K_1$ in the interaction rates, which leads to an exponential suppression for large $x_\varphi$. 
It becomes exponentially difficult to produce $\varphi$ particles at lower temperatures, and thus their abundance is suppressed. The continued rate for $\alpha\varphi\to\alpha\varphi$ falls off slower because there is a large number of pNGBs $\alpha$ relative to the number of scalars $\varphi$, which are able to transfer energy to the $\varphi$ and keep them in kinetic equilibrium. Initial rates do not suffer Boltzmann suppression, since they do not assume that the $\phi$ sector equilibrates. The $\phi$ and $\alpha$ number density is given by the number of Higgs decays/scatterings, which eventually plateaus. While the $\varphi \to \alpha \alpha$ initial rate keeps growing, the others decrease (and the $\alpha \alpha \to \varphi \varphi$ rate vanishes) at large $x$ because the particles lose energy.

\begin{figure}
    \centering
    \includegraphics[width=0.95\textwidth]{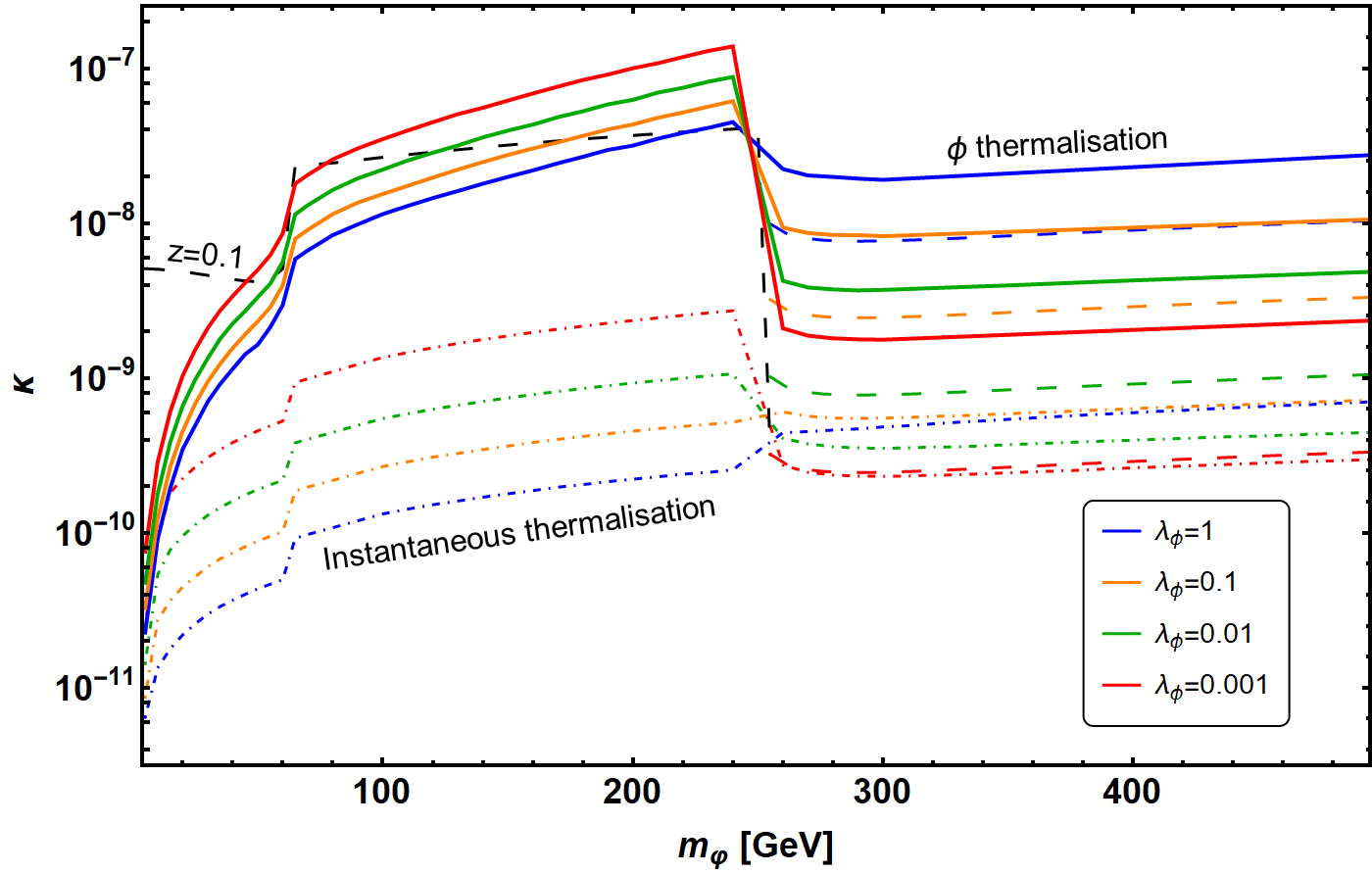}
    \caption{Portal coupling $\kappa$ required for $\phi$ sector thermalisation until at least $80\%$ of the $\phi$ have been produced (solid contours), assuming instantaneous thermalisation is possible (i.e. one instant of all relaxation rates larger being larger than the Hubble rate is sufficient, dotted contours), and for $z=0.1$ at late times (dashed contours). 
    For $m_\varphi \leq 2m_h$, the value of $z$ is $\lambda_\phi$-independent, however for $m_\varphi > 2m_h$ it depends on $\lambda_\phi$ via the $hh \to \varphi$ process. The observed dark matter abundance requires $z\simeq 0.1$, therefore above the dashed contours the sterile neutrino abundance would be too large.}
    \label{PhiThermPlot}
\end{figure}

In our numerical study, we consider thermal equilibrium to be reached if there is a period of time when the conditions for initial thermalisation hold, directly followed by a period of time when the conditions for continued thermalisation hold, until at least 80\% of the $\phi$ have been produced. This is satisfied above the solid contours in Fig.~\ref{PhiThermPlot}.
By contrast, the dotted contours indicate the values of $\kappa$ required for instantaneous thermalisation, i.e.~that there is at least one time during $\phi$ production that initial thermalisation and then continued thermalisation was achieved. 
It is notable that this is a significantly weaker condition. Fig.~\ref{PhiThermPlot} shows that for instantaneous thermalisation, values of $\kappa$ that are smaller by 1-2 orders of magnitude are allowed compared to those which enable sustained thermalisation during the period of dominant $\phi$ production. 
The dashed contours denote where $z=0.1$ is reached at late times, which indicates the relevant region of parameter space where the observed dark matter abundance is obtained.
For $m_\varphi \leq 2m_h$, this is independent of $\lambda_\phi$ and is represented by a black line, while for $m_\varphi > 2m_h$, it depends on $\lambda_\phi$ through the process $hh \to \varphi$, as can be seen from Eqn.~\eqref{eq:Mhhphi}.

For light $\varphi$ scalars with $m_\varphi<m_h/2$ the $\phi$ sector is efficiently heated via Higgs decays to a pair of $\varphi$ scalars and thus thermalisation is reached even for small portal couplings $\kappa\sim 10^{-11} - 10^{-10}$.
When $h\to\varphi\varphi$ becomes unavailable, the portal coupling $\kappa$ has to be substantially larger to heat the $\phi$ sector and thermalisation is only achieved for larger $\kappa\sim 10^{-8}$. 
For large $\varphi$ masses, $m_\varphi>2m_h$, although inverse Higgs decays $hh\to\varphi$ efficiently heat up the $\phi$ sector, the production of dark sector particles generally extends beyond the time when the dark sector drops out of thermal equilibrium and thus the resulting $\phi$ sector distribution does not follow a Bose-Einstein distribution. 
We find that for smaller $\lambda_\phi$, smaller values of $\kappa$ lead to thermalisation because the interaction rate of $hh\to\varphi$ scales like $\lambda_\phi^{-1}$, again see Eqn.~\eqref{eq:Mhhphi}. 
This implies that $T_\phi$ is larger for smaller values of $\lambda_\phi$, which counteracts the increase of the relaxation rates for increasing $\lambda_\phi$. Overall, we observe that thermalisation generally requires $\kappa\gtrsim 10^{-9}-10^{-8}$.

While thermalisation is achieved above the solid contours, the observed dark matter abundance is obtained close to the dashed contours. The analysis demonstrates that the dark matter will not follow a thermal distribution for $m_\varphi>2m_h$ irrespective of the value of $\lambda_\phi$. For $m_\varphi<m_h/2$, where Higgs decay $h\to\varphi\varphi$ is allowed, the value of $\kappa$ required to achieve $z=0.1$ is large enough to equilibrate the $\phi$ sector even for $\lambda_\phi\simeq 10^{-3}$. For $m_h/2<m_\varphi<2m_h$, the values of $\kappa$ to achieve $z=0.1$ are close to the thermalisation boundary and thus whether thermalisation is achieved or not depends on the quartic coupling $\lambda_\phi$.  

\section{Thermalisation of sterile neutrinos}
\label{sec:thermalisationN}

Having outlined above the conditions for the $\phi$ sector to thermalise, we now turn to the conditions for sterile neutrino equilibration with the $\phi$ bath. The $\phi-N$ interactions depend on the Yukawa coupling, $f$, cf.~Eqn.~\eqref{lagrangian}. 
Since we assume that $f = \sqrt{2}m_N/v_\phi \ll 1$ while $\lambda_\phi$ can be $\mathcal{O}(1)$, the timescales of thermalisation are different. 
We can therefore consider it to be a two-step process in which the sterile neutrinos only thermalise after the $\phi$ sector has already reached equilibrium. Thus, it is safe to entirely neglect the range of $T_\phi > T_{\phi,c}$ and to focus on temperatures $T_\phi$ below the critical dark sector temperature.

If the rates of $\varphi \leftrightarrow NN$, $\alpha \alpha \leftrightarrow NN$ and $\alpha N \to \alpha N$ are larger than the Hubble rate, then the sterile neutrinos thermalise and join the $\phi$ bath. 
In fact, it is sufficient for only one of $\varphi\leftrightarrow NN$ or $\alpha\alpha\leftrightarrow NN$ to be in equilibrium, as long as the $\phi$ sector remains in thermal equilibrium. 
The key conditions for thermalisation are therefore either of
\begin{align}
    &\langle \Gamma(\varphi \leftrightarrow NN) \rangle_{N} > H \, ,&
    &\langle \Gamma(\alpha \alpha \leftrightarrow NN ) \rangle_{N} > H \, .&
\end{align}
The process $\alpha N \to \alpha N $ also ensures kinetic equilibrium of $N$ for $\langle \Gamma(\alpha N \to \alpha N  ) \rangle_{N} > H$, however $\varphi \leftrightarrow NN$ and $\alpha \alpha \leftrightarrow NN$ are sufficient to achieve both chemical and kinetic equilibrium. Note that in principle a second possibility is that $N$ enters its own thermal bath via $NN \to NN$ scatterings, even if it does not equilibrate with the $\phi$ sector. 
The relevant decay widths and cross sections are
\begin{align}
    \Gamma(\varphi\to NN)  & = \frac{\lambda_\phi m_N^2}{8\pi m_\varphi} \\
    \sigma(NN\to \alpha\alpha)&\simeq
    \lim\limits_{\epsilon\to 0}
\frac{\lambda_\phi^2 m_N^2 s^2}{16 \pi  m_\varphi^4  } \frac{1}{(s-m_\varphi^2)^2 + m_\varphi^2 \Gamma_\varphi^2} 
\label{sigmannaa}
 \\\nonumber &\qquad\qquad
 \times
	 \left(1- \frac{ 2 m_\varphi^2\Gamma_\varphi^2 \epsilon^3 }{(s-m_\varphi^2)^2 + \epsilon^2 m_\varphi^2 \Gamma_\varphi^2} \frac{(s-m_\varphi^2)^2+m_\varphi^2\Gamma_\varphi^2}{(s-m_\varphi^2)^2+\epsilon^2m_\varphi^2\Gamma_\varphi^2}  \right)
     \\
    \sigma(\alpha\alpha\to NN) & =4\sigma(NN\to\alpha\alpha)
    \left(1-\tfrac{4m_N^2}{s}\right)
    \\
   	\sigma(\alpha N \to \alpha N) & \simeq 
  \frac{\lambda_\phi^2 m_N^2}{4\pi  m_\varphi^4} \left(\frac{s^2-3sm_\varphi^2-6m_\varphi^4}{2s(s+m_\varphi^2)} + \frac{3m_\varphi^4}{s^2}\ln \left(1+\frac{s}{m_\varphi^2}\right)\right)
    \label{sigmaanancs} 
    \\
    \label{sigmannnn}
    \sigma(NN\to NN)    &\simeq \lim_{\epsilon\to 0}  \frac{\lambda_\phi^2 m_N^4}{16\pi sm_\varphi^2} \Bigg[ \frac{3m_\varphi^4-3m_\varphi^2 s +2s^2+\Gamma_\varphi^2 (s+ 3m_\varphi^2) }{((s-m_\varphi^2)^2+m_\varphi^2\Gamma_\varphi^2)(s+m_\varphi^2)} 
    \\\nonumber&\qquad\qquad
    + \frac{4m_\varphi^2}{s}\frac{m_\varphi^4 +m_\varphi^2 s -2s^2+\Gamma_\varphi^2(s+m_\varphi^2)}{((s-m_\varphi^2)^2+m_\varphi^2\Gamma_\varphi^2)(s+2m_\varphi^2)}  \ln\left(1+\frac{s}{m_\varphi^2}\right) 
    \\\nonumber&\qquad\qquad
    - \frac{ 2s^2 \Gamma_\varphi^2  \epsilon^3}{((s-m_\varphi^2)^2+\epsilon^2m_\varphi^2\Gamma_\varphi^2)^2}
    \Bigg] \;. 
\end{align}
For each, we go to leading order in $m_N$, which is justified for $s\gg m_N^2$. 
For $s\ll m_\varphi^2$, the scattering cross section for $NN\to NN$ can be approximated by 
\begin{equation}
 \sigma(NN\to NN) \simeq  \frac{5 \lambda_\phi^2}{16\pi s} \frac{m_N^4}{m_\varphi^4} \, .
\end{equation}

The processes $NN \leftrightarrow \alpha \alpha$ and $NN\leftrightarrow NN$ both involve an $s$-channel diagram with a virtual $\varphi$, and therefore we need to perform real intermediate state subtraction. 
It is convenient to use the representation of the delta function, 
\begin{equation}
     \delta(s-m_\varphi^2) = \frac{2}{\pi} \lim_{\epsilon\to0} \frac{m_\varphi^3\Gamma_\varphi^3 \epsilon^3}{ ((s-m_\varphi^2)^2+\epsilon^2 m_\varphi^2 \Gamma_\varphi^2)^2} \, , 
\end{equation}
see Appendix A of \cite{Giudice:2003jh}. 
This is shown in Eqns.~\eqref{sigmannaa} and \eqref{sigmannnn}. 
For numerical evaluations it suffices to enforce that $\epsilon \ll \Gamma_\varphi/m_\varphi$. We neglect the $t$- and $u$-channel diagrams with a virtual $N$ in the calculation, because they are suppressed by an additional factor of $f$ compared to the $s$-channel contribution.

For $\alpha N \to \alpha N$, there are three tree-level diagrams: 
an $s$-channel diagram with a virtual $N$, a $u$-channel diagram with a virtual $N$, and a $t$-channel diagram with a virtual $\varphi$. 
However, we can neglect the $s$- and $u$-channel diagrams in the calculation because they are suppressed by an additional factor of $f$ compared to the $s$-channel diagram.

As in our study of the $\phi$ equilibration, we consider thermalisation in two stages. 
Firstly, in order to determine under what conditions the sterile neutrinos thermalise in the first place, we will assume that they initially have a delta function distribution and find whether they enter into equilibrium. 
Secondly, assuming that equilibration is achieved, we compute the thermally-averaged rates to find for how long they remain in equilibrium.

\subsection{Initial sterile neutrino thermalisation}
Consider the initial conditions that sterile neutrinos have a delta function distribution while the $\phi$ sector has already equilibrated. 
The rates $\langle \Gamma(\varphi \to NN) \rangle_{\varphi}$ and $\langle \Gamma(\alpha \alpha \to NN)\rangle_{\alpha}$ are insensitive to the $N$ distribution, up to negligible Pauli-blocking $(1-f_N)$ factors in the thermal averaging, therefore we postpone discussion of those to the following subsection. 
Here we address the processes $NN \to \varphi$, $NN \to \alpha \alpha$, $\alpha N \to \alpha N$ and $NN\to NN$.

As for the $\phi$ sector, we describe the initial sterile neutrino distribution by 
\begin{equation}
    f_N = C_N \delta(E_N - X_N) \, .
\end{equation}
The $N$ may be produced either by $\varphi \to NN$ or by $\alpha \alpha \to NN$. 
Since $\alpha \alpha \to NN$ dominates at the earliest times, let us take 
$X_N = \langle p_\alpha \rangle \simeq 2.7 T_\phi$. 
An estimate for the $N$ number density is $n_N \sim n_\varphi \langle \Gamma(\varphi \to NN) \rangle_{\varphi} /H + n_\alpha \langle \Gamma(\alpha \alpha \to NN) \rangle_{\alpha} /H$.
The number density from the distribution function is
\begin{equation}
n_N = \frac{C_N X_N^2}{2\pi^2} = 0.37 C_N T_\phi^2 \, ,
\end{equation}
neglecting $m_N \ll T_\phi$. 
We compute $C_N(x_\varphi)$ numerically by equating this to the estimate for $n_N$ from $\varphi$ decay and $\alpha$ scattering. 
The thermally averaged rates in the limit $m_N\ll T_\phi$ are
\begin{align}
    \langle\Gamma(NN \to \varphi) \rangle_{N} &\simeq \frac{2 C_N^2 m_\varphi \Gamma(\varphi\to NN)}{\pi^2 n_N}\label{Ntherm1} \\
    \langle \Gamma(NN \to \alpha \alpha) \rangle_{N} &\simeq \frac{C_N^2 m_\varphi^4}{16\pi^4x_\varphi^4 n_N} \int_0^{4\chi_N^2} dr \, \sigma(s=T_\phi^2 r)\, r
    \label{Ntherm2}\\
    \langle \Gamma(\alpha N \to \alpha N) \rangle_{N} &\simeq
    \frac{C_N m_\varphi^5}{16\pi^4 x_\varphi^5 n_{N} } \int_0^\infty dr ~\sigma(s=T_\phi^2 r)\, r \left( \frac{r}{4\chi_N} - \ln(e^{r/4\chi_N} - 1) \right) 
    \label{Ntherm3}\\
        \langle \Gamma(NN \to NN) \rangle_{N} &
        \simeq\frac{C_N^2 m_\varphi^4}{16 \pi^4x_\varphi^4 n_N} \int_0^{4\chi_N^2} dr \, \sigma(s=T_\phi^2 r)\, r \, ,
\end{align}
neglecting Bose-enhancement and Pauli-blocking factors, with $\chi_N\equiv X_N/T_\phi$ and $r\equiv s/T_\phi^2$.

\subsection{Continued sterile neutrino thermalisation}
The thermally averaged rates when $\varphi$, $\alpha$ and $N$ have MB, BE and FD distributions, respectively, are 
    \begin{align}
    \langle \Gamma(\varphi \leftrightarrow NN) \rangle_{N} &\simeq \frac{\Gamma(\varphi\to NN) m_\varphi^3 }{4 \pi^2 x_\varphi^2 n_N} \int_{x_\varphi}^{\infty} \frac{dF_\varphi}{e^{F_\varphi} +1} \ln\left[\frac{\cosh\left(\frac{F_\varphi + \sqrt{F_\varphi^2 - x_\varphi^2}}{4}\right)}{\cosh\left(\frac{F_\varphi - \sqrt{F_\varphi^2 - x_\varphi^2}}{4}\right)} \right]  \\
\langle \Gamma( NN \to \alpha \alpha)  \rangle_N &\simeq
\frac{  m_\varphi^6 }{8 \pi^4 x_\varphi^6 n_N}
 \int_{0}^\infty dr \int_{\sqrt{r}}^\infty dF_+  \frac{\sigma(s=T_\phi^2 r) r}{e^{F_+}-1}\ln \left[ \frac{\cosh\left(\tfrac{F_+}{4}\left[1+ \sqrt{1-\tfrac{r}{F_+^2}}\right] \right)}
    {  \cosh\left(\tfrac{F_+}{4}\left[1- \sqrt{1-\tfrac{r}{F_+^2}}\right]\right)  } \right]  \label{initNNaa} \\
    \langle \Gamma (\alpha N\to \alpha N)\rangle_{N} 
          & \simeq \frac{ m_\varphi^6}{16 \pi^4 x_\varphi^6 n_N} \int_{0}^\infty dr \int_{\sqrt{r}}^\infty dF_+ \frac{ \sigma(s=T_\phi^2 r) r  }{e^{F_+}+1} \ln\left[\frac{\sinh\left(\frac{F_+ + \sqrt{F_+^2-r}}{2}\right)}
          {\sinh\left(\frac{F_+ - \sqrt{F_+^2-r}}{2}\right)}
          \right]   
    \label{anancs}\\
    \langle \Gamma( NN \to NN)  \rangle_N & \simeq  
\frac{  m_\varphi^6 }{32 \pi^4 x_\varphi^6 n_N}
 \int_{0}^\infty dr \int_{\sqrt{r}}^\infty dF_+  \frac{\sigma(s=T_\phi^2 r) r}{e^{F_+}-1}\ln \left[ \frac{\cosh\left(\tfrac{F_+}{4}\left[1+ \sqrt{1-\tfrac{r}{F_+^2}}\right] \right)}
    {  \cosh\left(\tfrac{F_+}{4}\left[1- \sqrt{1-\tfrac{r}{F_+^2}}\right]\right)  } \right] \,,  
    \end{align}
again to leading order in $m_N$. To be conservative, we neglected the $(1+f_{\varphi, \alpha})$ Bose-enhancement factors in the rates, and for the Pauli-blocking terms in $\alpha N \to \alpha N$ and $NN \to NN$ we used the property that $(1-f_N) \geq 1/2$. We do not separately impose the condition $\langle\Gamma(\alpha\alpha \to NN)\rangle_N >H$, since it does not pose any new constraint for $4m_N^2\ll T_\phi^2$. 
The interaction rate $NN\to NN$ turns out to be tiny, because the cross section is proportional to $m_N^4$ and thus very suppressed. We numerically verified that the rate does not reach equilibrium even for the largest sterile neutrino mass and smallest scalar mass that we consider in the analysis, $m_N = 100$ keV and $m_\varphi = 1$ GeV. Consequently, we neglect it in the following discussion.
Combining the above with Eqns.~\eqref{Ntherm1}, \eqref{Ntherm2}, and \eqref{Ntherm3}, we can find for how long thermalised sterile neutrinos remain in equilibrium.

\subsection{Discussion}
\begin{figure}
    \centering
    \includegraphics[width=0.8\linewidth]{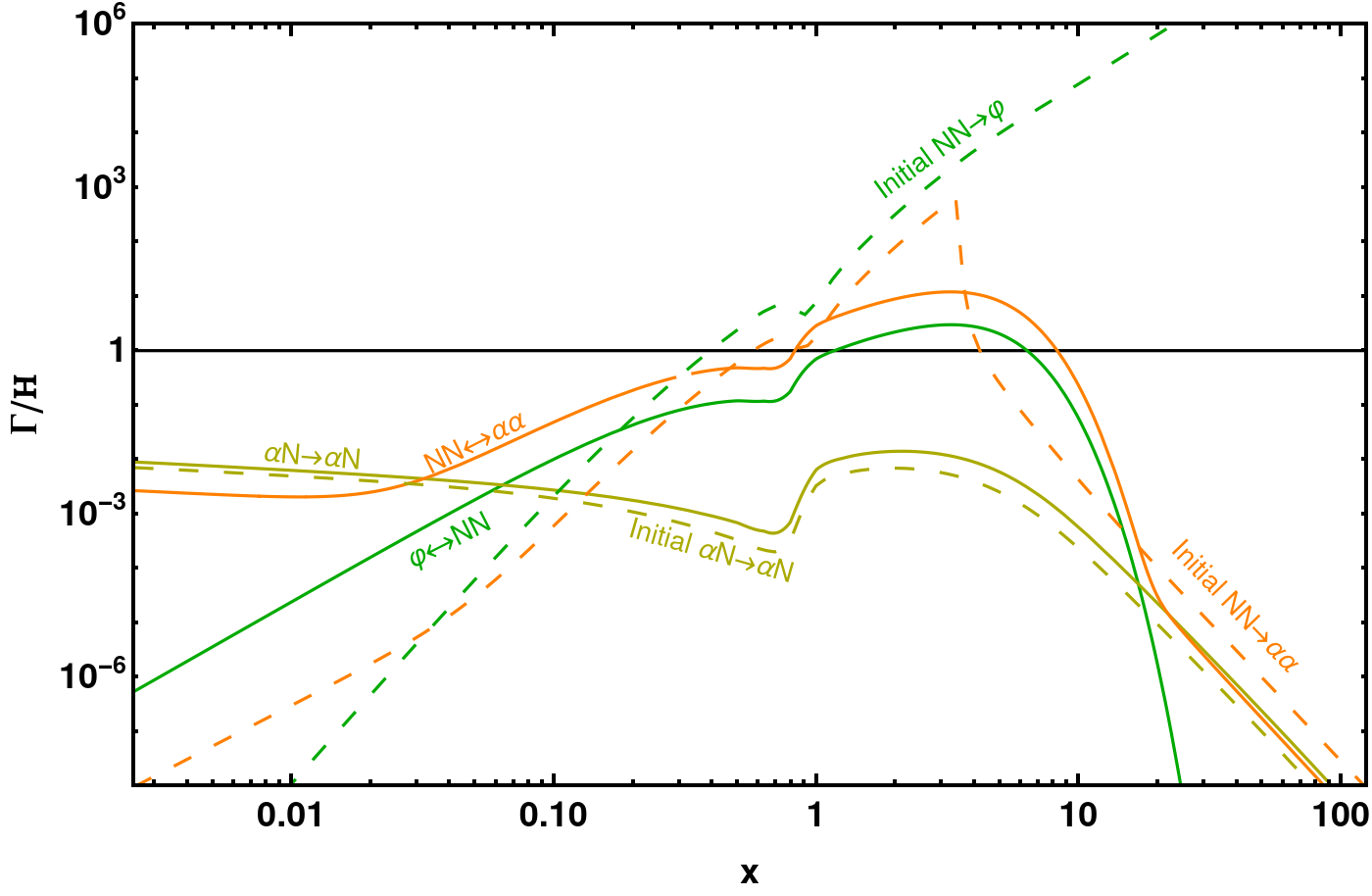}
    \caption{Relaxation rates of the sterile neutrinos for $\lambda_\phi = 0.1$, $\kappa = 10^{-8}$, $m_\varphi = 50$ GeV and $m_N = 80$ keV.
    Continued (initial) thermalisation rates are shown as solid (dashed) lines. 
    }
    \label{fig:Nrates}
\end{figure}

In Fig.~\ref{fig:Nrates} we illustrate the relaxation rates relative to the Hubble rate as a function of $x$. The continued $\varphi \leftrightarrow NN$ relaxation rate increases with $x$ until the $\varphi$ abundance becomes Boltzmann suppressed at $x\gtrsim z m_h/m_\phi$.
For large $T_\phi$ (small $x$), the scattering rates do not drop off as quickly as the decay rate, because they do not suffer from a suppression with a Lorentz factor $m_\varphi/E_\varphi$ like the decay rate. Similarly, for large $x$ the decrease of the scattering rates is softer because the pNGB $\alpha$ is light and thus its distribution function not exponentially suppressed.
The slight drop and subsequent increase in the rates around $x\sim 0.78$ originates from the electroweak phase transition: the scalar $\varphi$ briefly becomes non-relativistic with $T_\phi<m_\varphi$, but the increased energy injection from Higgs decays increases the temperature and consequently the rates, cf. the thermalisation of the $\phi$ sector. 
The drop in the initial $NN\to \alpha\alpha$ relaxation rate occurs due to the behaviour of the resonance at $x_\varphi=2 X_N$. 
The resonance region provides the largest contribution for small $x$, but becomes inaccessible for large $x$.

The abundance and average energy of the $N$ in the initial thermalisation regime are computed from decays and annihilations of the $\varphi$ and $\alpha$. At late times, these are suppressed, as indeed we see in Fig.~\ref{fig:Nrates}. From then on, the $N$ number density and average energy simply scale with the expansion, just like a thermal distribution of relativistic particles. 
Thus, for large $x$ the initial relaxation rate shows the same scaling as the continued relaxation rate.
Finally, both the continued and the initial relaxation rates for the $\alpha N \to \alpha N$ scattering show the same behaviour which is expected, because we assumed that $\alpha$ is in thermal equilibrium. 
Following Fig.~\ref{PhiRatesPlot}, the $\phi$ sector remains in chemical equilibrium until $x\sim 12$ and in kinetic equilibrium for $x\geq 100$ and therefore $\alpha$ follows a Bose-Einstein distribution even when the $N$ relaxation rates fall out of equilibrium at $x\sim 10$. 

\begin{figure}
    \centering
    \includegraphics[width=0.8\linewidth]{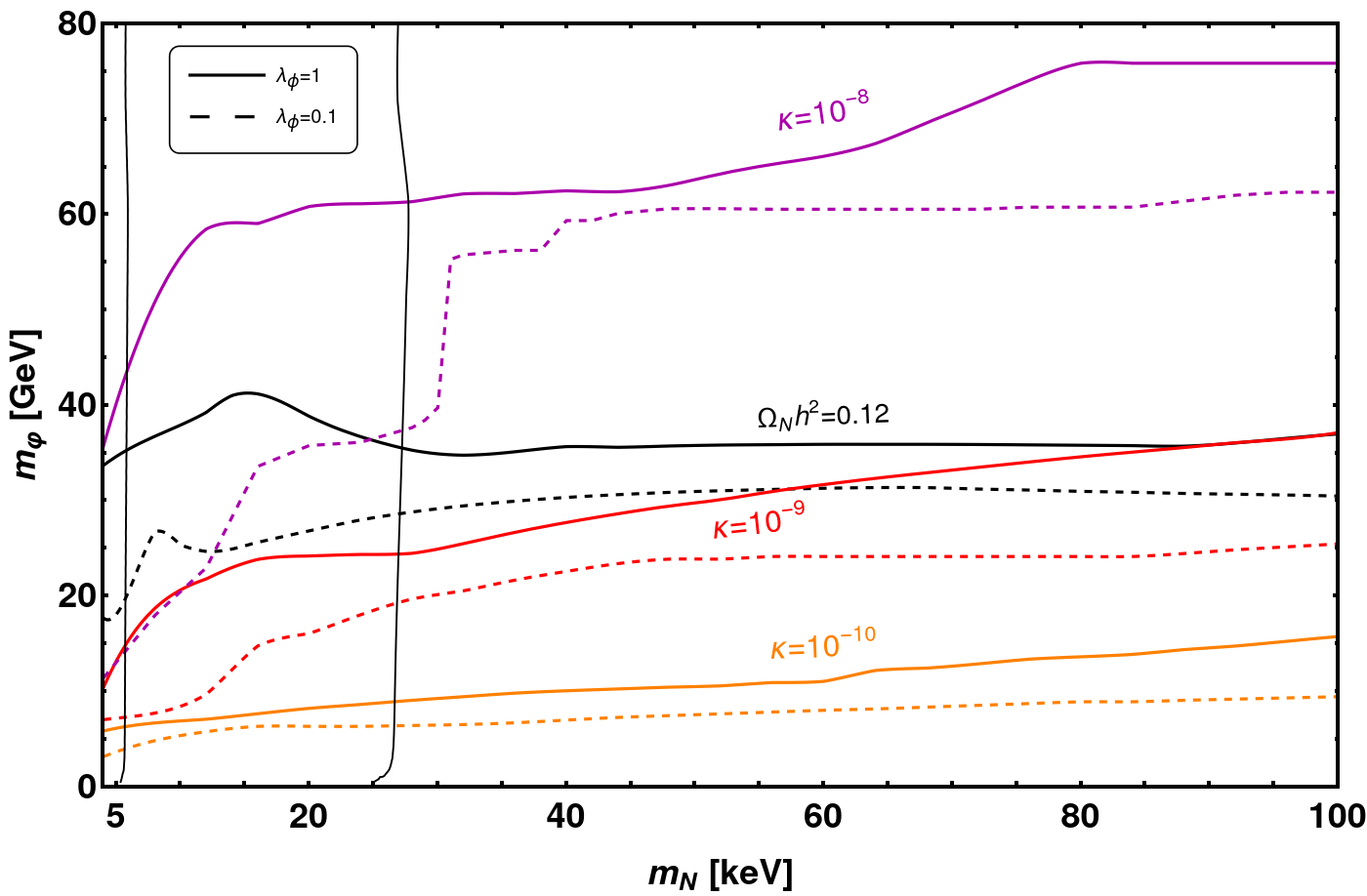}
    \caption{Thermalisation of the sterile neutrinos. Thermalisation is achieved for values of $m_\varphi$ below the contours. The coloured contours correspond to the given values of $\kappa$, the black contours correspond to the value of $\kappa$ required for $\Omega_N h^2 = 0.12$, which is itself a function of $m_\varphi$ and $m_N$. The contours of fixed free-streaming horizon are given for the correct relic abundance. }
    \label{fig:Nthermalisation}
\end{figure}

In Fig.~\ref{fig:Nthermalisation} we illustrate the thermalisation as a function of the masses, for different values of $\kappa$ and $\lambda_\phi$. Thermalisation is achieved below the different contours. Solid contours are for $\lambda_\phi=1$ and dashed contours for $\lambda_\phi=0.1$. The purple, red and orange contours correspond to $\kappa=10^{-8}$, $\kappa=10^{-9}$ and $\kappa=10^{-10}$, respectively. 
For $\lambda_\phi=0.1$ (dashed contours) the temperature for which thermalisation is achieved is close to the electroweak phase transition and thus the thermalisation is very sensitive to the input parameters. This becomes particularly obvious for $\kappa=10^{-8}$ (dashed purple contour). 
For the black contours, $\kappa$ is fixed by $\Omega_N h^2=0.12$. 
A comparison between the black and the colored curves shows that the correct dark matter abundance requires values of $\kappa$ that decreases from $\kappa \simeq 10^{-8}$ to $\kappa \simeq 10^{-9}$ as $m_N$ increases for $\lambda_\phi = 1$, and from a few times $10^{-8}$ to a few times $10^{-9}$ for $\lambda_\phi = 0.1$. 
The decreasing $\kappa$ as $m_N$ increases, for fixed relic abundance, is largely due to the relation $\Omega_N \propto \kappa^{3/2} m_N$ discussed in Sec.~\ref{CalcTphi}. 
For $m_N< 100$ keV, we find that thermalisation of the sterile neutrinos requires reasonably light $m_\varphi$. Higgs decay is therefore the dominant production mode for most of the parameter space where sterile neutrinos thermalise. 
Note also that these values of $\kappa$ and $m_\varphi$ are consistent with thermalisation of the $\phi$ sector, cf.~Fig.~\ref{PhiThermPlot}. 
The free-streaming horizons with $\lambda_{\rm FS}=10 (66)$ kpc are indicated by two almost vertical black contours in Fig.~\ref{fig:Nthermalisation}. The free-streaming horizon depends on the sterile neutrino mass $m_N$ and the temperature ratio $z$, as shown in Eqn.~\eqref{lambdafs}. 
The slight deviation from exactly straight contours originates from the $z$ dependence on $m_\varphi$, particularly around $m_\varphi\simeq m_h/2$, above which Higgs decay becomes inaccessible as a production mechanism for dark sector particles.

%%%%%%%%%%%%%%%%%%%%%%%%%%%%%%%%%%%%%%%%%%%%%%
%%%%%%%%%%%%%%%%%%%%%%%%%%%%%%%%%%%%%%%%%%%%%%

\section{Conclusions}
\label{sec:conclusion}

We have investigated the production of keV sterile neutrino dark matter via a frozen-in complex scalar field. In contrast to the scenario with a real scalar field, the dark sector may thermalise and thus result in a drastically different phenomenology. After the complex scalar field $\phi$ develops a non-vanishing vacuum expectation value $v_\phi$, the sterile neutrinos becomes massive and lepton number is broken.  The dark sector thus consists of the sterile neutrino $N$, the CP even scalar $\varphi$ and the pNGB $\alpha$. Depending on the mass of the $\varphi$, the dominant production mechanism of $\varphi$ is from Higgs decay, scattering or inverse decay.
Demanding that the keV sterile neutrino lifetime exceeds the age of the universe requires tiny active-sterile neutrino mixing, which thus has been neglected in the analysis. This leads to an enhanced symmetry and the absence of DM production from active-sterile neutrino oscillations. 

Subsequently, the pNGB $\alpha$ is produced via $\varphi$ decay and $\varphi\varphi \to \alpha\alpha$ scattering. These processes, together with $\alpha\varphi\to \alpha\varphi$ and $\alpha\alpha\to\alpha\alpha$, also ensure that the dark scalar sector thermalises for a large region of parameter space, depending on the Higgs portal coupling $\kappa$ and the quartic dark sector coupling $\lambda_\phi$. 
Its temperature, $T_\phi$, is smaller than the temperature of the SM thermal bath, $T$. 
Reproducing the observed DM abundance requires a dark sector temperature ratio $z=T_\phi/T \simeq 0.1$: this puts an upper limit on the mass of the thermalised CP even scalar $\varphi$, as shown in Fig. \ref{PhiThermPlot}. 
For $m_\varphi<m_h/2$ the dark sector with $z\simeq 0.1$ thermalises easily, even for $\lambda_\phi\sim 10^{-3}$. 
For heavier $\varphi$ scalar masses, $m_h/2<m_\varphi<2m_h$, the required temperature $z\simeq 0.1$ is close to the thermalisation boundary, indeed whether the sector thermalises depends sensitively on the value of $\lambda_\phi$. 
Finally, for $m_\varphi>2m_h$ dark matter will not follow a thermal distribution irrespective of $\lambda_\phi$, assuming the correct relic abundance.

The sterile neutrinos $N$ are produced predominantly via $\varphi\to NN$ decays and $\alpha\alpha \to NN$ scattering. 
The thermalisation of the sterile neutrinos is more difficult to achieve than in the $\phi$ sector, because the sterile neutrino Yukawa coupling $f$ is small by construction. 
We find that the thermalisation of this sector generally requires small $\varphi$ masses and large scalar quartic coupling $\lambda_\phi$. For $m_N<100$ keV, all of the parameter space in which thermalisation is achieved and the correct relic abundance is produced corresponds to the CP even scalar mass being $m_\varphi < m_h/2$, see Fig.~\ref{fig:Nthermalisation}. 
In this case, although the pNGBs and sterile neutrinos modify the effective number of neutrinos in the early Universe, $N_{\rm eff}$, their contribution is below the sensitivity of CMB-S4~\cite{CMB-S4:2016ple}, being $\mathcal{O}(10^{-3})$.

Despite the thermalisation of the $\phi$ sector and the sterile neutrinos, the influence of the latter on structure formation in this model is similar to other sterile neutrino DM models. 
Its free-streaming horizon depends mainly on its mass, $m_N$, and the temperature ratio $z$, depending only on $m_\varphi$ via $z$. It agrees with the bound on early-decoupled fermionic DM~\cite{Irsic:2017ixq}, for which there is a lower limit $m_\text{DM} > 5.3$ keV, because the combination of the sterile neutrino mass and the explanation of the observed dark matter abundance fixes the temperature ratio and thus the free-streaming horizon.

In this work, we only considered the case where the dark sector fully thermalises. We would like to stress that the dark sector may not thermalise or only partially thermalise, and thus there is a rich phenomenology to explore which we leave for future work. Thermalisation is not essential to explain the correct dark matter abundance. 
As illustrated in Fig.~\ref{numberplot}, the dark sector particle abundances will generally be smaller if they do not thermalise than if they had, therefore a lack of thermalisation could be compensated for by a slightly larger value of $\kappa$ to produce the right amount of DM.
Moreover, we only considered one sterile neutrino $N$, and thus the $U(1)$ symmetry is anomalous. It is straightforward to make the $U(1)$ symmetry anomaly free by introducing three sterile neutrinos. The additional sterile neutrinos could generate neutrino masses like in the neutrino minimal Standard Model~\cite{Asaka:2005an}. 

%%%%%%%%%%%%%%%%%%%%%%%%%%%%%%%%%%%%%%%%%%%%
\begin{acknowledgements} 
We thank Michele Frigerio, Karsten Jedamzik, Archil Kobakhidze, Oleg Sushkov, and Matthew O'Brien for illuminating discussions. 
R.C. thanks the UNSW School of Physics for its hospitality during part of this project. 
This project has received support from the IISN convention 4.4503.15 and the Australian Research Council through the ARC Discovery Project DP200101470. 
\end{acknowledgements}
%%%%%%%%%%%%%%%%%%%%%%%%%%%%%%%%%%%%%%%%%%%%%
%%%%%%%%%%%%%%%%%%%%%%%%%%%%%%%%%%%%%%%%%%%%%

\appendix
\appendixpage
\addappheadtotoc

\section{Self energy of the pNGB}
\label{app:selfenergy}
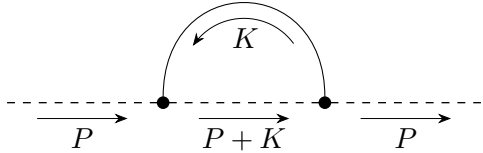
\begin{figure}[b]
\centering
 \begin{tikzpicture}[node distance=2cm]
    \begin{feynhand}
      \vertex (a1);
        \vertex[right=of a1,dot] (v1) {};
      \vertex[right=of v1,dot] (v4) {};
      \vertex[right=of v4] (a2);
      \propag [scalar,mom'={$P$}] (a1) to (v1);
      \propag [scalar,mom'={$P+K$}] (v1) to (v4);
      \propag [scalar,mom'={$P$}] (v4) to (a2);
      \propag [out=90,in=90,mom={$K$},looseness=2] (v4) to (v1);
    \end{feynhand}
  \end{tikzpicture}
  
  \caption{Relevant contributions to the imaginary part of the self-energy at 1-loop order. Solid lines indicate $\varphi$ scalar propagators and dashed lines $\alpha$ pNGB propagators.}
  \label{fig:self-energy}
\end{figure}

We follow the Kobes-Semenoff~\cite{Kobes:1985kc,Kobes:1986za} formalism to calculate the imaginary part of the pNGB self energy in a thermal background. The relevant 1-loop self-energy diagram is shown in Fig.~\ref{fig:self-energy}. Note that the 1-loop diagram with the quartic interaction does not contribute to the imaginary part of the self-energy. 
%%%%%%%%%%%%%%%%%%%%%%%%%%%%%%%%%%%%%%%%%%%%%%%%%%%%%%%%%%%%%%%%%5
We denote the 4-momentum of the pNGB by $P^\mu$, its energy by $p_0$ and the absolute magnitude of its 3-momentum by $p$. In the result we have to distinguish between three cases depending on $P^2$.
For a real on-shell pNGB with $P^2=0$ we find
\begin{align}
  \mathrm{Im}\bar\Pi(P^2=0)
&=\epsilon(p_0)\frac{m_\varphi^4}{16 \pi v_\phi^2 \beta p} \left(
\beta p-\ln\left|\frac{f(\frac{m_\varphi^2}{4p})}{f(p+\frac{m_\varphi^2}{4p})}\right|
\right) \, ,
\end{align}
where $\beta \equiv 1/T$, $f$ is the distribution function, and $\epsilon$ is the sign function. For space-like $P^2$ with $P^2<0$ we find
\begin{align}
  \mathrm{Im}\bar\Pi(P) & = \frac{\epsilon(p_0)}{16 \pi v_\phi^2 \beta p} \left[P^2 - m_\varphi^2\right]^2 \theta(-P^2)\left[
  2\beta p -
  \ln\left|\frac{f(\frac{p_0-p}{2}\frac{m_\varphi^2}{P^2})}{f(p+\frac{p_0-p}{2}\frac{m_\varphi^2}{P^2})}\right|
  -
  \ln\left|\frac{f(-\frac{p_0+p}{2}\frac{m_\varphi^2}{P^2})}{f(p-\frac{p_0+p}{2}\frac{m_\varphi^2}{P^2})}\right|
  \right]
\end{align}
Finally for time-like $P^2$ with $P^2>0$ we have to distinguish whether the energy $p_0$ is positive or negative. Note that we slightly modified the equations above in order to obtain a more compact expression by introducing absolute magnitudes. We find
\begin{align}
  \mathrm{Im}\bar\Pi(P) & = \frac{\epsilon(p_0)}{16 \pi v_\phi^2 \beta p} \left[P^2 - m_\varphi^2\right]^2 \theta(P^2)
  \\\nonumber&
  \Bigg\{
  \theta(P^2-m_\varphi^2)
    \left[\ln\left|\frac{f(\frac{p_0+p}{2}+\frac{(p_0-p)}{2}\frac{m_\varphi^2}{P^2})}{f(-\frac{p_0-p}{2}+\frac{p_0-p}{2}\frac{m_\varphi^2}{P^2})}\right|
-
\ln\left|\frac{f(\frac{p_0-p}{2}+\frac{p_0+p}{2}\frac{m_\varphi^2}{P^2})}{f(-\frac{p_0+p}{2}+\frac{p_0+p}{2}\frac{m_\varphi^2}{P^2})}\right|
\right]
  \\\nonumber&
  +\theta(m_\varphi^2-P^2)
  \Bigg[
  \ln\left|\frac{f\left(-\frac{p_0+p}{2}+\frac{|p_0|+p}{2}\frac{m_\varphi^2}{P^2}\right)}{f\left(\frac{p_0-p}{2}+\frac{|p_0|+p}{2}\frac{m_\varphi^2}{P^2}\right)}\right|
  \\\nonumber&\qquad
  -\theta(p_0-p)\ln\left|\frac{f\left(-\frac{p_0-p}{2}+\frac{p_0-p}{2}\frac{m_\varphi^2}{P^2}\right)}{f\left(\frac{p_0+p}{2}+\frac{p_0-p}{2}\frac{m_\varphi^2}{P^2}\right)}\right|
  -\theta(-p-p_0)
  \ln\left|\frac{f\left(\frac{\sqrt{p^2+m_\varphi^2}-p_0}{2}\right)}{f\left(\frac{\sqrt{p^2+m_\varphi^2}+p_0}{2}\right)}\right|
  \Bigg]
  \Bigg\}
\end{align}
Using the above, we obtain for the propagator of the pNGB $\alpha$,
\begin{equation}
D_\alpha(P) = \frac{i}{P^2 - i \mathrm{Im}\bar\Pi(P) +i\epsilon}
\;.
\end{equation}

\section{\texorpdfstring{$\alpha \varphi\to \alpha\varphi$}{alpha phi -> alpha phi} scattering}
\label{app:aPaP}

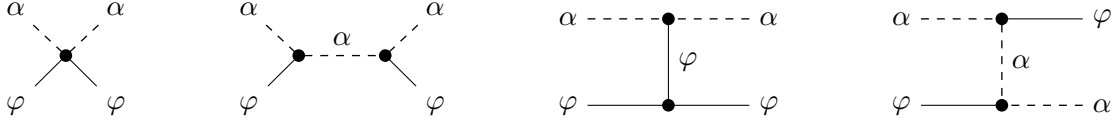
\begin{figure}
    \centering
    \begin{tikzpicture}
    \begin{feynhand}
      \vertex[dot] (v) {};
        \vertex[above left=0.5 of v] (i1) {$\alpha$};
        \vertex[below left=0.5 of v] (i2) {$\varphi$};
        \vertex[above right=0.5 of v] (f1) {$\alpha$};
        \vertex[below right=0.5 of v] (f2) {$\varphi$};
      \propag [scalar] (i1) to (v);
      \propag (i2) to (v);
      \propag [scalar] (f1) to (v);
      \propag  (f2) to (v);
    \end{feynhand}
    \end{tikzpicture}
    \hspace{1cm}
    \begin{tikzpicture}
    \begin{feynhand}
      \vertex[dot] (v1) {};
      \vertex[dot,right=of v1] (v2) {};
        \vertex[above left=0.5 of v1] (i1) {$\alpha$};
        \vertex[below left=0.5 of v1] (i2) {$\varphi$};
        \vertex[above right=0.5 of v2] (f1) {$\alpha$};
        \vertex[below right=0.5 of v2] (f2) {$\varphi$};
      \propag [scalar] (i1) to (v1);
      \propag (i2) to (v1);
      \propag [scalar] (f1) to (v2);
      \propag  (f2) to (v2);
      \propag[scalar]  (v1) to[edge label=$\alpha$] (v2);
    \end{feynhand}
    \end{tikzpicture}
    \hspace{1cm}
  \begin{tikzpicture}
    \begin{feynhand}
      \vertex[dot] (v1) {};
      \vertex[dot,below=of v1] (v2) {};
        \vertex[left=of v1] (i1) {$\alpha$};
        \vertex[left=of v2] (i2) {$\varphi$};
        \vertex[right=of v1] (f1) {$\alpha$};
        \vertex[right=of v2] (f2) {$\varphi$};
      \propag [scalar] (i1) to (v1);
      \propag (i2) to (v2);
      \propag [scalar] (f1) to (v1);
      \propag  (f2) to (v2);
      \propag (v1) to[edge label=$\varphi$] (v2);
    \end{feynhand}
    \end{tikzpicture}
    \hspace{1cm}
\begin{tikzpicture}
    \begin{feynhand}
      \vertex[dot] (v1) {};
      \vertex[dot,below=of v1] (v2) {};
        \vertex[left=of v1] (i1) {$\alpha$};
        \vertex[left=of v2] (i2) {$\varphi$};
        \vertex[right=of v1] (f1) {$\varphi$};
        \vertex[right=of v2] (f2) {$\alpha$};
      \propag[scalar] (i1) to (v1);
      \propag (i2) to (v2);
      \propag (f1) to (v1);
      \propag[scalar]  (f2) to (v2);
      \propag[scalar] (v1) to[edge label=$\alpha$] (v2);
    \end{feynhand}
    \end{tikzpicture}
 
    \caption{Contributions to $\alpha\varphi\to \alpha\varphi$ scattering. Dashed (solid) lines correspond to $\alpha$ ($\varphi$) propagators. }
    \label{fig:ap2ap}
\end{figure}

We discuss the scattering of $\alpha\varphi\to \alpha\varphi$ in more detail to properly address the $u$-channel soft-collinear singularity. The relevant Feynman diagrams are given in Fig.~\ref{fig:ap2ap}.
We find for the matrix element
\begin{equation}
    i\mathcal{M} = -2i\lambda_\phi \left[\frac{t}{m_\varphi^2} + \frac{3t}{t-m_\varphi^2} + \frac{(s-m_\varphi^2)^2}{m_\varphi^2 s} + \frac{(u-m_\varphi^2)^2}{u - i\mathrm{Im}\bar \Pi}\right] \, ,
\end{equation}
in terms of the Mandelstam variables $s=(p_\alpha+p_\varphi)^2$, $t=(p_\alpha-p_\alpha^\prime)^2$, and $u=(p_\alpha-p_\varphi^\prime)^2$. Here $p_\alpha$ ($p_\varphi$) denotes the 4-momentum of the incoming $\alpha$ ($\varphi$) particles and $p_\alpha^\prime$ ($p_\varphi^\prime$) denotes the 4-momentum of the outgoing $\alpha$ ($\varphi$) particles. We included the correction from the thermal self energy of $\alpha$, derived in the previous appendix, for the $u$-channel diagram to regularise the divergence.\footnote{The contribution of thermal self energy corrections to the regularisation of $t$ (and $u$) channel diagrams in cosmology was recently discussed in Ref.~\cite{Grzadkowski:2021kgi}.} 
For all other internal propagators, we neglected it as there are no other singularities. We also neglected the $\varphi$ width for the $\varphi$ propagator. The self-energy complicates the evaluation of the interaction rate due to its explicit energy dependence. We thus split up the interaction rate into three parts: the $u$-channel diagram, the other three diagrams, and the interference term.

The $u$-independent term can be directly evaluated using standard techniques in the centre of mass frame and is given in Eqn.~\eqref{eq:sigma_aPaP}.
In order to evaluate the $u$-dependent part, we have to resort to an explicit calculation in a general frame. After evaluating the integral over $p_\varphi^\prime$ using the delta function, we introduce spherical coordinates and choose the following parameterisation for the 3-momenta,
\begin{align}
\mathbf{p_\varphi} & = p_\varphi (0,0,1)^T & 
\mathbf{p_\alpha} & = p_\alpha (0,s_x,c_x)^T &
\mathbf{p_\alpha^\prime} & = p_\alpha^\prime (s_\beta s_\theta,c_\beta s_\theta,c_\theta)^T \, ,
\end{align}
where $s_X\equiv \sin X$ and $c_X\equiv \cos X$. Approximating the distribution function by a Maxwell-Boltzmann distribution and neglecting statistical factors, we obtain for the interaction rate
\begin{align}
    \langle \Gamma_u \rangle_\varphi 
	&= \frac{\lambda_\phi^2 m_\varphi}{8(2\pi)^5 x_\varphi n_\varphi} \int_{m_\varphi}^\infty dE_\varphi \int_{-1}^1 d\cos\theta
	\left( \int_{-\infty}^{0} I_\alpha^- du +  \int_0^{m_\varphi^2} I_\alpha^+du  \right)
	\\\nonumber &\qquad\qquad\qquad\qquad\qquad\qquad
	\frac{(m_\varphi^2-u)^5 p_\varphi f(E_\varphi) }{m_\varphi^4|u-i\mathrm{Im}\Pi_\alpha|^2 (E_\varphi -p_\varphi \cos\theta)^2\sqrt{p_\alpha^{\prime2} + p_\varphi^2 -2p_\alpha^\prime p_\varphi \cos\theta}}
\end{align}
with $|\mathbf{p_\alpha^\prime}-\mathbf{p_\varphi}|=\sqrt{p_\alpha^{\prime 2} + p_\varphi^2 -2 p_\varphi p_\alpha^\prime c_\theta}$, the Mandelstam variable 
$u=m_\varphi^2-2 p_\alpha^\prime (E_\varphi - p_\varphi c_\theta)\leq m_\varphi^2$ and the two functions $I_\alpha^\mp$ which are non-zero for $u<0$ and $0<u<m_\varphi^2$, respectively,
\begin{align}
	I_{\alpha}^- & = 
\theta(-u)\Bigg[\theta(p_{\alpha-})\Big(
		e^{-p_{\alpha+}/T} \theta(p_{\alpha+}-p_\alpha^\prime+E_\varphi)
	+e^{-(p_\alpha^\prime-E_\varphi)/T} \theta(p_\alpha^\prime-E_\varphi -p_{\alpha+})
	 \nonumber\\&
+(\theta(E_\varphi-p_\alpha^\prime)  + e^{-(p_\alpha^\prime-E_\varphi)/T} \theta(p_\alpha^\prime-E_\varphi) - e^{-p_{\alpha+}/T}) \theta(p_{\alpha-} -p_\alpha^\prime-E_\varphi)
	\Big) \notag
	 \\\nonumber &
	+\theta(p_{\alpha+})\theta(-p_{\alpha-})
	\Big(
		e^{-p_{\alpha+}/T} \theta(p_{\alpha+} -p_\alpha^\prime+E_\varphi)
		+e^{-(p_\alpha^\prime-E_\varphi)/T} \theta(p_\alpha^\prime-E_\varphi-p_{\alpha+})
	\Big)
	 \\ &
	 +\theta(-p_{\alpha+})
	 \Big(
		 \theta(E_\varphi-p_\alpha^\prime) 
		 +e^{-(p_\alpha^\prime-E_\varphi)/T} \theta(p_\alpha^\prime-E_\varphi)
	 \Big)
\Bigg] 
 \\
 I_{\alpha}^{+} 
	&=\theta(u)\theta(m_\varphi^2-u)\theta(p_{\alpha-} -p_\alpha^\prime + E_\varphi)
	\Bigg[ \notag 
	\\\nonumber& 
		\theta(p_{\alpha+})
		\Big(
			e^{-p_{\alpha+}/T} \theta(p_{\alpha +} -p_\alpha^\prime+E_\varphi)
			+e^{-(p_\alpha^\prime-E_\varphi)/T}\theta(p_\alpha^\prime-E_\varphi -p_{\alpha+})
		\Big)
	\\ &
		+\theta(p_{\alpha-}) \theta(-p_{\alpha+})
		\Big(
			\theta(E_\varphi-p_\alpha^\prime) + e^{-(p_\alpha^\prime-E_\varphi)/T} \theta(p_\alpha^\prime -E_\varphi)
		\Big)
		-e^{-p_{\alpha-}/T}\theta(p_{\alpha-})
	\Bigg] \, ,
\end{align}
which are defined piecewise and depend on 
\begin{align}
    p_{\alpha,\pm} & = \frac{u-m_\varphi^2}{2u} \left[p_\alpha^\prime-E_\varphi \pm |\mathbf{p_\alpha^\prime}-\mathbf{p_\varphi}|\right]
    \;.
\end{align}

Similarly, we obtain for the relevant interaction rate for initial thermalisation using the distribution functions $f_{\varphi}(E_\varphi)=C_{\varphi}\delta(E_\varphi-X_\varphi)$ and $f_\alpha(p_\alpha) =C_\alpha\delta(p_\alpha-X_\alpha)$, 
\begin{align}
	  \langle \Gamma_u \rangle_{\varphi} 
	 = \frac{\lambda_\phi^2 C_\alpha C_\varphi}{8(2\pi)^5 n_\varphi} \int_{-1}^1 d\cos\theta
	&\left( \int_{-\infty}^{0} I_\alpha^- du +  \int_0^{m_\varphi^2} I_\alpha^+du  \right)
	\\\nonumber &
	\frac{(m_\varphi^2-u)^5 p_\varphi \theta(X_\varphi-m_\varphi) }{m_\varphi^4|u-i\mathrm{Im}\Pi_\alpha|^2 (X_\varphi -p_\varphi \cos\theta)^2\sqrt{p_\alpha^{\prime2} + p_\varphi^2 -2p_\alpha^\prime p_\varphi \cos\theta}} \, ,
\end{align}
with $p_\varphi \equiv\sqrt{X_\varphi^2-m_\varphi^2}$ and the two piecewise defined functions
\begin{align}
	I_{\alpha}^-  & =
\theta(-u)\Bigg[\theta(p_{\alpha-})\Big(
	\theta(X_\alpha-p_{\alpha+}) \theta(p_{\alpha+}-p_\alpha^\prime+X_\varphi)
+\theta(X_\alpha-(p_\alpha^\prime-X_\varphi)) \theta(p_\alpha^\prime-X_\varphi -p_{\alpha+})
	 \nonumber\\&
 +(\theta(X_\varphi-p_\alpha^\prime)  + \theta(X_\alpha-(p_\alpha^\prime-X_\varphi)) \theta(p_\alpha^\prime-X_\varphi) - \theta(p_{\alpha+}-X_\alpha)) \theta(p_{\alpha-} -p_\alpha^\prime-X_\varphi)
	\Big) \notag 
	 \\\nonumber &
	+\theta(p_{\alpha+})\theta(-p_{\alpha-})
	\Big(
	\theta(X_\alpha-p_{\alpha+}) \theta(p_{\alpha+} -p_\alpha^\prime+X_\varphi)
+\theta(X_\alpha-(p_\alpha^\prime-X_\varphi)) \theta(p_\alpha^\prime-X_\varphi-p_{\alpha+})
	\Big)
	 \\ &
	 +\theta(-p_{\alpha+})
	 \Big(
		 \theta(X_\varphi-p_\alpha^\prime)
		 +\theta(X_\alpha-(p_\alpha^\prime-X_\varphi)) \theta(p_\alpha^\prime-X_\varphi)
	 \Big)
\Bigg]
\;,
\end{align}
\begin{align}
I_{\alpha}^{+}
	=\theta(u)&\theta(m_\varphi^2-u)\theta(p_{\alpha-} -p_\alpha^\prime + X_\varphi)
	\Bigg[ \\&
		\theta(p_{\alpha+})
		\Big(
		\theta(X_\alpha-p_{\alpha+}) \theta(p_{\alpha +} -p_\alpha^\prime+X_\varphi)
	+ \theta(X_\alpha-(p_\alpha^\prime-X_\varphi))\theta(p_\alpha^\prime-X_\varphi -p_{\alpha+})
		\Big)
	\nonumber\\ &
		+\theta(p_{\alpha-}) \theta(-p_{\alpha+})
		\Big(
		\theta(X_\varphi-p_\alpha^\prime) + \theta(X_\alpha-(p_\alpha^\prime-X_\varphi)) \theta(p_\alpha^\prime -X_\varphi)
		\Big)
	-\theta(p_{\alpha-}-X_\alpha)\theta(p_{\alpha-})
	\Bigg] \notag 
	\;.
\end{align}
Using the Cuba library~\cite{Hahn:2004fe}, we numerically find that the $u$-channel contribution is generally subdominant for the parameter region of interest with a few exceptions, where it becomes of the same order of magnitude as the $u$-independent part. In the main part of the text we hence neglect the $u$-channel contribution and the interference term in order to reduce the computational complexity involved with evaluating the multi-dimensional integral for the $u$-dependent contributions.

%%%%%%%%%%%%%%%%%%%%%%%%%%%%%%%%%%%%%%%%%%%%%%%%%%%%%%%%%%%%%%%%%%%%%%%%%%%%%%%%
%%%%%%%%%%%%%%%%%%%%%%%%%%%%%%%%%%%%%%%%%%%%%%%%%%%%%%%%%%%%%%%%%%%%%%%%%%%%%%%%

\bibliography{keV.bib}

\end{document}